\documentclass{emulateapj}
\usepackage{apjfonts}
\slugcomment{To appear in {\it The Astrophysical Journal}.}
\shorttitle{A NEW SAMPLE OF BURIED AGNs SELECTED FROM THE 2XMM CATALOGUE}
\shortauthors{NOGUCHI, TERASHIMA, \& AWAKI}

\begin{document}
\title{A New Sample of Buried Active Galactic Nuclei Selected from the Second {\it XMM-Newton} Serendipitous Source Catalogue}
\author{Kazuhisa Noguchi, Yuichi Terashima, and Hisamitsu Awaki}
\affil{Department of Physics, Ehime University, Matsuyama, Ehime 790-8577, Japan}

\begin{abstract}
We present the results of X-ray spectral analysis of 22 active
galactic nuclei (AGNs) with a small scattering fraction selected from
the Second {\it XMM-Newton} Serendipitous Source Catalogue using
hardness ratios. They are candidates of buried AGNs, since a scattering
fraction, which is a fraction of scattered emission by the
circumnuclear photoionized gas with respect to direct emission, can be
used to estimate the size of the opening part of an obscuring
torus. Their X-ray spectra are modeled by a combination of a power
law with a photon index of 1.5$-$2 absorbed by a column density of
$\sim$ 10$^{23-24}$ cm$^{-2}$, an unabsorbed power law, narrow
Gaussian lines, and some additional soft components. We find that
scattering fractions of 20 among 22 objects are less than a typical
value ($\sim$ 3\%) for Seyfert2s observed so far. In particular, those
of eight objects are smaller than 0.5\%, which are in the range for
buried AGNs found in recent hard X-ray surveys. Moreover, [\ion{O}{3}]
$\lambda$5007 luminosities at given X-ray luminosities for some
objects are smaller than those for Seyfert2s previously known. This
fact could be interpreted as a smaller size of optical narrow emission
line regions produced in the opening direction of the obscuring
torus. These results indicate that they are strong candidates for the AGN
buried in a very geometrically thick torus.
\end{abstract}

\section{Introduction}
It is widely accepted that the cosmic X-ray background (CXB) is produced
by the integrated emission of faint extragalactic pointlike sources
(Brandt \& Hasinger 2005). {\it XMM-Newton} and {\it Chandra} resolved
80\%$-$100\% of the CXB at $<$ 2 keV, while the resolved fraction of
the CXB at hard X-rays (8$-$12 keV) decreased to only $\approx$ 50\%
(Worsley et al. 2005). Various observations indicate that a large
fraction of active galactic nuclei (AGNs) is hidden by a large amount
of cold material (e.g., Awaki et al. 1991; Comastri 2004). According to population
synthesis models of the CXB (Comastri et al. 1995; Ueda et al. 2003;
Gilli et al. 2007), the peak intensity of the CXB spectrum at 30 keV can
be explained by considering contribution of hidden AGNs. Such a
population is yet to be understood, because the direct emission from the
nucleus is absorbed by surrounding cold gas and is hard to be fully
explored with X-ray observations below 10 keV. Hard X-ray surveys
performed with {\it Swift}/BAT (15$-$200 keV; Markwardt et al. 2005;
Tueller et al. 2008) and {\it INTEGRAL} (10$-$100 keV; Bassani et
al. 2006; Beckmann et al. 2006; Sazonov et al. 2007) are suitable for
unveiling such a type of AGNs with much less selection biases than
surveys at lower energies. In fact, AGNs buried in a large amount of
matter have been discovered by {\it Suzaku} follow up observations of
a sample selected by the {\it Swift/}BAT survey (Ueda et al. 2007).

In a unified model of an AGN (e.g., Antonucci 1993), torus-like gas is
surrounding a supermassive black hole (SMBH), and photoionized gas is
created in the opening part of the torus by radiation from the
nucleus. If an AGN is observed from the torus side, absorbed direct
emission and scattered light by the photoionized gas will be observed in
an X-ray spectrum. The fraction of scattered light to direct emission
(scattering fraction) can be used to estimate the opening angle of the
torus. The fractions for AGNs found by Ueda et al (2007) are extremely
small ($<$ 0.5\%), whereas a typical value is $\sim$ 3\% (Turner et al. 1997;
Bianchi and Guainazzi 2007). Furthermore, Winter et al (2008) found a
similar type of AGNs by {\it XMM-Newton} observations of {\it
  Swift/}BAT selected AGNs. They would be buried in a geometrically thick
torus with a very small opening angle assuming that the scattering
fraction reflects the solid angle of the opening part of the torus. In
an early stage of the evolution of galaxies and their central black
holes, a large amount of gas responsible for active star formation may
be closely related to obscuration of the nucleus. Therefore, AGNs
almost fully covered by an absorber are an important class of objects
in studying evolution of AGNs and their hosts.  Moreover, they might be
significant contributors to the CXB at hard X-rays. Testing a selection
technique to find such AGNs and understanding the properties of the 
population are of significant interest for exploring these issues.

We search for buried AGNs with a scattering fraction of 0.5\% or less
using the Second {\it XMM-Newton} Serendipitous Source Catalogue
(2XMM) and the archival data of {\it XMM-Newton}. We selected
candidate sources from the catalogue using hardness ratios (HRs) and 
scattering fractions calculated by analyzing spectra obtained with
{\it XMM-Newton}. The selection method of candidate sources is
described in Section 2. Our results of spectral analysis are presented
in Section 3 and their characteristics are discussed in Section
4. Section 5 summarizes our conclusions. We adopt ({\it H}$_{\rm 0}$,
$\Omega_{\rm m}$, $\Omega_{\rm \lambda}$) $=$ (70 km
s$^{-1}$Mpc$^{-1}$, 0.3, 0.7) throughout this paper.

\section{Selection of Candidates for a Buried AGN}

Our sample was selected from the 2XMM Catalogue that has been
assembled by the {\it XMM-Newton} Survey Science Centre (Watson et
al. 2009). This catalogue contains 246897 X-ray source detections. The
median flux in the full energy band (0.2$-$12 keV) is $\sim$ 2.5
$\times$ 10$^{-14}$ erg cm$^{-2}$ s$^{-1}$ and about 20\% of the
sources have total fluxes below 1 $\times$ 10$^{-14}$ erg cm$^{-2}$
s$^{-1}$.

We used HR3 and HR4 among some HRs defined in the 2XMM Catalogue to
select candidates of buried AGNs. These HRs are defined as
\begin{displaymath}
 {\rm HR}3=\frac{{\rm CR}(2.0-4.5\ {\rm keV})-{\rm CR}(1.0-2.0\ {\rm keV})}{{\rm CR}(2.0-4.5\ {\rm keV})+{\rm CR}(1.0-2.0\ {\rm keV})} 
\end{displaymath}
and 
\begin{displaymath}
 {\rm HR}4=\frac{{\rm CR}(4.5-12\ {\rm keV})-{\rm CR}(2.0-4.5\ {\rm keV})}{{\rm CR}(4.5-12\ {\rm keV})+{\rm CR}(2.0-4.5\ {\rm keV})} , 
\end{displaymath}
where CR(1.0$-$2.0 keV), CR(2.0$-$4.5 keV), and CR(4.5$-$12 keV) are
count rates in the 1.0$-$2.0, 2.0$-$4.5, and 4.5$-$12 keV bands,
respectively. The values of HR given in the 2XMM Catalogue were
calculated using count rates measured by the {\tt emldetect} task in
the {\it XMM-Newton} Science Analysis System (SAS). If direct emission
from an AGN is absorbed by cold material with {\it N}$_{\rm H}$ $\sim$
10$^{23}$ cm$^{-2}$, CR(1.0$-$2.0 keV) and CR(2.0$-$4.5 keV) are
dominated by the soft component such as scattered emission and
absorbed direct emission, respectively. In the case of {\it N}$_{\rm
  H}$ $\sim$ 10$^{24}$ cm$^{-2}$, CR(2.0$-$4.5 keV) and CR(4.5$-$12
keV) are dominated by the soft component and direct emission,
respectively. Therefore, HR3 and HR4 can be used to efficiently select
objects with {\it N}$_{\rm H}$ $\sim$ 10$^{23}$ cm$^{-2}$ and
10$^{24}$ cm$^{-2}$, respectively.

In the selection process, we required sources to satisfy the following
conditions: 1) count rate for EPIC-pn in 0.2$-$12 keV $>$ 0.05 cts
s$^{-1}$, 2) high Galactic latitude ($|b|$ $>$ 20$^\circ$), and 3)
error of HRs $\leq$ 0.2 at a 90\% confidence level. The
errors of HRs shown in the 2XMM Catalogue were derived from count
rates measured by the SAS task {\tt emldetect}. 4627 sources among
246897 satisfied these criteria.

We simulated AGN spectra in XSPEC (version 11.2) to calculate HRs
expected for an object with a low scattering fraction, using the
response function of the EPIC-pn. The spectral model assumed in the
simulation is a combination of absorbed and unabsorbed power laws,
which correspond to direct and scattered components, respectively. We
fixed the photon indices of both power-law components at 1.9, which is
a typical value of Seyfert 2 galaxies (e.g., Smith \& Done 1996). The
scattering fraction is defined as a ratio between the normalizations
of the two power laws. We simulated spectra for scattering fractions
of 10, 5, 3, 1, and 0.5\%, and log{\it N}$_{\rm H}$ from 20.5 to 24.5
cm$^{-2}$ at a logarithmic step of 0.1. The expected HRs are shown in
Figure 1 as solid and dashed lines. The five solid lines represent the
scattering fractions of 10, 5, 3, 1, and 0.5\% from inside to
outside. The three dashed lines correspond to objects with log{\it
  N}$_{\rm H}$ = 23, 23.5, and 24 from lower right to upper left. 4627
sources, which fulfilled the conditions defined above, are also
plotted as crosses. Objects with a small scattering fraction are
located in the upper right portion in this figure.

In order to select buried AGNs with a scattering fraction of $<$ 0.5\%
as many as possible, we selected 23 objects located upper right of the
line for a scattering fraction of 1\% since there are uncertainties in
HR values and all the spectra may not be explained by the simple model
defined above. The 23 candidates are plotted with circles in Figure
1. In Table 1, we list the source name, AGN type, redshift, Galactic
column density toward the source, start date of the observation,
exposure time after data screening given in Section 3, and count rates
in the 0.4$-$10 keV band. References for the AGN type are also shown
in Table 1. Redshifts are taken from NASA/IPAC Extragalactic Database
(NED). The criteria adopted in this paper to select a buried AGN hold at
low redshift only, because we assumed $z$ = 0 in the simulation of AGN
spectra used to calculate the expected HRs. In fact, most
of AGNs in our sample are at $z$ $<$ 0.1. The Galactic column
densities are calculated from 21 cm measurements (Kalberla et al. 2005)
using the {\tt nh} tool at the NASA's High Energy Astrophysics Science
Archive Research Center. {\it XMM-Newton} results of some of the objects in
our sample have been published. References for them are listed in
Table 1. Since 2XMM J234349.7$-$151700 is a star, we excluded it from
our sample for spectral analysis in Section 3.

\begin{figure}[!htb]
\centering
\includegraphics[angle=270,scale=.40]{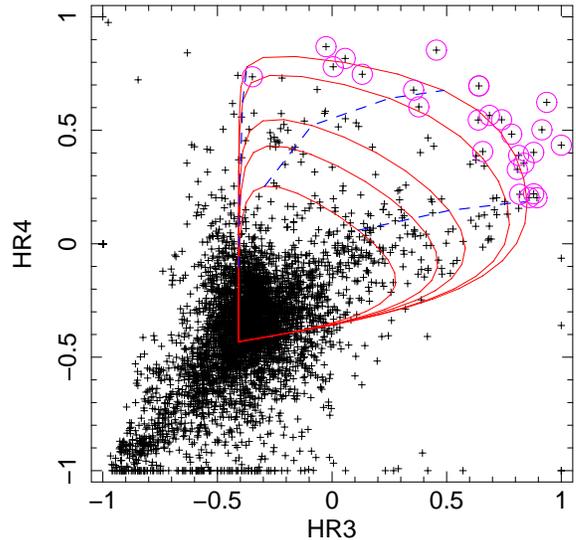}
\caption{Distribution of HR for the 2XMM Catalogue sources
  (Crosses). These are satisfied all conditions; count rate in 0.2-12
  keV $>$ 0.05 cts s$^{-1}$, $|b|$ $>$ 20, and HR error $\leq$
  0.2. Our sample is shown by circles. Solid lines show the HRs
  expected for the scattering fraction of 10, 5, 3, 1, and
  0.5\% from inside to outside. Dashed lines correspond to log$N$$_{\rm H}$ of
  23, 23.5, and 24 cm$^{-2}$ from lower right to upper left.}
\end{figure}

\section{Data Reduction and Analysis}
In order to calculate accurate values of the scattering fraction for
our sample, we analyzed their X-ray spectra obtained with EPIC-pn. The
data were reduced with the SAS version 7.1. We created calibrated
photon event files for the EPIC-pn camera from the observation data
files (ODF). The ancillary response files and detector response
matrices were generated using the {\tt arfgen} and {\tt rmfgen} tasks,
respectively. X-ray events corresponding to patterns 0$-$4 were
selected from the event files. We extracted source spectra from
circular regions with a radius in a range
10$^{\prime}$$^{\prime}$$-$40$^{\prime}$$^{\prime}$, depending on the
brightness of the source. Background spectra were taken from a region
near the target. Time intervals with very high background rates were
identified in light curves above 10 keV and discarded. We fit spectra
of the sample in the 0.4$-$10 keV range with various models by using
XSPEC version 11.2. All spectra except one were binned so that each
spectral bin contains more than 20 counts per bin to enable usage of
the $\chi^2$ fit statistics. Since the total number of counts for one
object, 2MASX J12544196$-$3019224, is small, the same way of binning
resulted is very small number of bins. We therefore used $C$-statistic
(Cash 1979) to fit the unbinned spectrum of this object. The quoted
errors correspond to a 90\% confidence level for one interesting
parameter (i.e., $\Delta\chi^2$ or $\Delta$$C$ = 2.71).

{\tabletypesize{\scriptsize}
\begin{deluxetable*}{llcccccccc}
\setlength{\tabcolsep}{-3pt}
\tablecolumns{10}
\tablecaption{Properties of Sample}
\tablewidth{0pt}
%\tablewidth{\columnwidth}
%\setlength{\tabcolsep}{0.035in}
%\renewcommand{\arraystretch}{1.05}
\tablehead{
 \colhead{2XMM Name}  &  
 \colhead{Other Name}  & 
 \colhead{Class$^a$}  &  
 \colhead{Reference$^b$}  &
 \colhead{Redshift$^c$}  & 
 \colhead{$N_{\rm H}$$^d$} & 
 \colhead{Start Date}  & 
 \colhead{Exposure$^e$}  & 
 \colhead{Count Rate$^f$} & 
 \colhead{Reference$^g$}\\
 \colhead{}   & 
 \colhead{}   &
 \colhead{}   &  
 \colhead{}   & 
 \colhead{}   & 
 \colhead{($10^{20}$ cm$^{-2}$)} &    
 \colhead{}    & 
 \colhead{(s)} & 
 \colhead{(counts s$^{-1}$)}&
 \colhead{} 
 }
 \startdata
  2XMM J004847.1+315724     & Mrk 348                   &  Sy2    & 1       & 0.015034     & 5.79 & 2002 Jul 18  & 19323   & 2.0    & 6,7 \\
  2XMM J010852.8+132015     & 3C 33                     &  Sy2    & 2       & 0.059700     & 3.23 & 2004 Jan 21  & 6291    & 0.19   & 8,9 \\
  2XMM J022813.4$-$031502   & 2MASX J02281350$-$0315023 & \nodata & \nodata & 0.078706$^h$ & 2.21 & 2002 Jul 26  & 8487    & 0.040  & \nodata \\
  2XMM J025512.1$-$001100   & NGC 1142                  &  Sy2    & 1       & 0.028847     & 5.81 & 2006 Jan 28  & 8879    & 0.23   & 10 \\
  2XMM J035854.4+102602     & 3C 98                     &  Sy2    & 1       & 0.030454     & 10.2 & 2002 Sep 07  & 13771   & 0.22   & 11 \\
  2XMM J090053.8+385616     & B2 0857+39                &  Sy2    & 3       & 0.229000     & 2.05 & 2004 Apr 13  & 20151   & 0.013  & \nodata \\ 
  2XMM J091957.9+371127     & IC 2461                   & \nodata & \nodata & 0.007539     & 1.08 & 2003 Apr 15  & 48567   & 0.19   & \nodata \\
  2XMM J103352.6+004403     & 2MASX J10335255+0044033   & Sy1.8   & 1       & 0.131500     & 4.81 & 2005 Dec 11  & 4391    & 0.048  & \nodata \\ 
  2XMM J113543.9+490216     & MCG+08$-$21$-$065         & \nodata & \nodata & 0.029687     & 1.75 & 2003 Nov 24  & 15007   & 0.029  & \nodata \\
  2XMM J120429.7+201858     & NGC 4074                  &  Sy2    & 1       & 0.022445     & 2.30 & 2003 Jan 02  & 2767    & 0.11   & \nodata \\
  2XMM J120929.7+434106     & NGC 4138                  & Sy1.9   & 4       & 0.002962     & 1.25 & 2001 Nov 26  & 8801    & 0.45   & 12,13 \\
  2XMM J122546.7+123943     & NGC 4388                  & Sy1.9   & 4       & 0.008419     & 2.58 & 2002 Dec 12  & 7346    & 1.5    & 14 \\
  2XMM J123536.6$-$395433   & NGC 4507                  &  Sy2    & 1       & 0.011801     & 7.04 & 2001 Jan 04  & 34020   & 0.99   & 15 \\
  2XMM J123854.6$-$271827   & ESO 506$-$G027            &  Sy2    & 5       & 0.025024     & 5.45 & 2006 Jan 24  & 8276    & 0.22   & 10 \\
  2XMM J125442.6$-$301928   & 2MASX J12544196$-$3019224 & \nodata & \nodata & 0.056105     & 5.99 & 2004 Jan 15  & 10314   & 0.013  & \nodata \\ 
  2XMM J130414.3$-$102021   & NGC 4939                  &  Sy2    & 1       & 0.010374     & 3.30 & 2002 Jan 03  & 11503   & 0.050  & 16 \\
  2XMM J133326.1$-$340052   & ESO 383$-$G18             & Sy1.8   & 1       & 0.012412     & 4.04 & 2006 Jan 10  & 12573   & 0.43   & \nodata \\
  2XMM J183820.3$-$652539   & ESO 103$-$G035            &  Sy2    & 1       & 0.013286     & 5.71 & 2002 Mar 15  & 8392    & 1.6    & 17,18 \\
  2XMM J201958.9$-$523718   & IC 4995                   &  Sy2    & 1       & 0.016094     & 4.08 & 2004 Sep 25  & 6630    & 0.068  & 19 \\
  2XMM J213147.2$-$425051   & NGC 7070A                 & \nodata & \nodata & 0.007976     & 2.56 & 2004 Oct 28  & 26377   & 0.15   & 20 \\
  2XMM J220201.8$-$315210   & NGC 7172                  &  Sy2    & 1       & 0.008683     & 1.95 & 2007 Apr 24  & 27833   & 3.4    & 6 \\
  2XMM J223603.5+335833     & NGC 7319                  &  Sy2    & 1       & 0.022507     & 6.15 & 2001 Dec 07  & 34010   & 0.11   & 21 \\
  2XMM J234349.7$-$151700   & R Aqr                     & \nodata & \nodata & \nodata      & 1.87 & 2005 Jul 30  & \nodata & \nodata & \nodata    
\enddata     
\tablenotetext{a}{Optical classification. ``Sy2'' represents Seyfert 2. ``Sy1.8'' and ``Sy1.9'' represent intermediate Seyfert.}
\tablenotetext{b}{References for optical classification.}
\tablenotetext{c}{Redshift are taken from the NED, expect for 2MASX J02281350$-$0315023.}
\tablenotetext{d}{Galactic column density by 21 cm measurement (Kalberla et al. 2005).}
\tablenotetext{e}{Cleaned exposure of EPIC-pn.}
\tablenotetext{f}{Count rate in the 0.4$-$10 keV band.}
\tablenotetext{g}{References for published {\it XMM-Newton} results.}
\tablenotetext{h}{Determined by an Fe$-$K$\alpha$ emission line in the X-ray spectrum analyzed in this paper. The line center energy is assumed to be 6.4 keV in the source rest frame.}\\
\tablerefs{(1)Veron-Cetty \& Veron 2006; (2)Fanaroff \& Riley 1974; (3)Kiuchi et al 2006; (4) Ho et al. 1997; (5)Tueller et al 2008; (6) Awaki et al. 2006; (7) Guainazzi \& Bianchi 2007; (8) Evans et al. 2006; (9) Kraft et al 2007; (10) Winter et al. 2008; (11) Isobe et al. 2005; (12) Foschini et al. 2002; (13) Cappi et al. 2006; (14) Beckmann et al. 2004; (15) Matt et al. 2004; (16) Guainazzi et al 2005a; (17) Shinozaki et al 2006; (18) De Rosa et al 2008; (19) Guainazzi et al. 2005b; (20) Trinchieri et al. 2008; (21) Trinchieri et al. 2005.}
\end{deluxetable*}
}

\clearpage

\subsection{Baseline Model}
We fitted spectra of the 22 sources with a model consisting of two
power laws and a narrow Gaussian line to account for an Fe K emission
at 6.4 keV, all modified by Galactic absorption using {\tt phabs} in
XSPEC. An absorption by cold matter at the redshift of the source
({\tt zphabs} in XSPEC) was added to the Gaussian and one of the power
laws.  We assumed that the photon indices of both power laws are
same. Hereafter, we refer to this model as the baseline model.  Of 22
sources, spectra of seven sources (2MASX J02281350$-$0315023, B2
0857+39, 2MASX J10335255+0044033, MCG +08$-$21$-$065, NGC 4074, 2MASX
J12544196$-$3019224, and NGC 7070A) are well fitted with the baseline
model. In these fits, the photon indices ($\Gamma$) were fixed at 1.9
except for NGC 7070A, since uncertainties of $\Gamma$ became large
($>$ 20\%) if $\Gamma$ was left free.
%(The errors are quoted at the 90\% confidence level for one interesting parameter ($\Delta$$C$ = 2.71)). 
The results are shown in Table
\ref{table:baseline}. The best-fit $N_{\rm H}$ values for the seven
objects are about 2$\times$10$^{23}$ cm$^{-2}$.  The spectra of these
objects are shown in Figure 2.

{\tabletypesize{\scriptsize}
\begin{deluxetable*}{lccccccccc}
\tablecolumns{10}
\tablecaption{Spectral Parameters for the Baseline Model}
\setlength{\tabcolsep}{-3pt}
\tablewidth{0pt}
\tablehead{
\colhead{}\\    
 \colhead{Name}            & \colhead{$N_{\rm H}$}        & 
 \colhead{$\Gamma$}        & \colhead{$A_{\rm int}$$^a$}           & 
 \colhead{$E_{\rm line}$}   & \colhead{$\sigma$}          &  
 \colhead{EW}              &\colhead{$A_{\rm ga}$$^b$}    &
 \colhead{$A_{\rm scat}$$^c$}   &  \colhead{$\chi^2_{\rm \nu}$(dof)}        \\ 
 \colhead{}                &\colhead{($10^{22}$ cm$^{-2}$)}&
 \colhead{}                &                             &
 \colhead{(keV)}           & \colhead{(eV)}              & 
 \colhead{(eV)}            &                            &  
                           & \\
}               
 \startdata
 Mrk 348$^d$              & 13.2  & 1.56   & 10.4 & 6.42 & 40 & 54& 3.16 & 6.26 & 2.18(150) \\
 3C 33$^d$                & 35.4  & 1.33 & 0.931 & 6.44 & 100 & 260 & 2.17 & 3.17 & 1.52(42)  \\         
 2MASX J02281350$-$0315023& 20.6$^{+4.6}_{-3.5}$    & 1.9(f)                 & 0.75$\pm0.14$              & 6.40(f)               & 10(f)                & 170($<$360)     & 0.41($<$0.84)          & 0.35$\pm0.17$           & 0.72(14)  \\ 
 NGC 1142                 & 63.9$\pm3.2$          & 2.32$^{+0.06}_{-0.12}$    & 16.20$\pm0.28$             &6.410$\pm0.040$          & 0($<$60)             &180$^{+60}_{-50}$ & 4.0$\pm1.3$            & 4.32$\pm0.33$           & 1.32(92)  \\ 
 3C 98                    & 14.43$^{+0.98}_{-0.89}$ & 1.900$^{+0.065}_{-0.069}$ & 1.889$\pm0.068$           & 6.40(f)                & 10(f)              &63$\pm49$        & 0.36$\pm0.28$          & 2.23$\pm0.26$           & 1.17(143)  \\  
 B2 0857+39               & 19.0$^{+5.5}_{-3.9}$    & 1.9(f)                  & 0.41$^{+0.10}_{-0.09}$     & 6.40(f)                & 10(f)                &42($<$240)       & 0.06($<$0.36)           & 0.25$\pm0.16$           & 1.16(15)   \\  
 IC 2461                  & 7.05$\pm0.20$         & 1.728$^{+0.055}_{-0.053}$  & 1.400$^{+0.037}_{-0.042}$   & 6.409$^{+0.036}_{-0.035}$ & 106$^{+49}_{-36}$   &$210^{+54}_{-48}$ & 1.21$^{+0.31}_{-0.27}$    & 0.323$^{+0.065}_{-0.064}$ & 0.96(118) \\ 
 2MASX J10335255+0044033   & 19.5$^{+5.9}_{-4.2}$    & 1.9(f)                  & 0.52$^{+0.13}_{-0.10}$     & 6.40(f)                & 10(f)               & 0($<$105)      & 0($<$0.18)            & 0.73$^{+0.20}_{-0.27}$    & 0.83(8)  \\
 MCG +08$-$21$-$065       & 23.6$^{+3.9}_{-3.2}$   & 1.9(f)                  & 0.81$^{+0.14}_{-0.12}$     &6.40(f)                & 10(f)                &190$\pm140$     & 0.46$^{+0.34}_{-0.35}$    & 0.21$\pm0.16$           & 0.76(23)   \\  
 NGC 4074                 & 19.2$^{+4.4}_{-3.5}$    & 1.9(f)                  & 2.27$^{+0.58}_{-0.46}$      & 6.40(f)                & 10(f)               &190($<$430)     & 1.3($<$2.9)             & 3.12$\pm0.73$           & 0.98(13)  \\
 NGC 4138                 & 7.66$^{+0.45}_{-0.40}$  & 1.592$^{+0.058}_{-0.059}$  & 1.769$^{+0.051}_{-0.054}$   &6.380$^{+0.061}_{-0.040}$ &  10(f)               &85$^{+55}_{-45}$  & 0.82$^{+0.43}_{-0.45}$    & 1.75$^{+0.24}_{-0.22}$    & 1.11(180)\\ 
 NGC 4388$^d$             & 32.8                  & 1.88                     & 23.0                  & 6.43                  & 38                  &170              & 12.1                   & 11.7                    & 2.18(162)  \\  
 NGC 4507$^d$             & 52.6                  & 2.44                     & 62.3                     &6.41                    & 0                   &120              & 8.06                    & 20.8                    & 5.12(210)   \\ 
 ESO 506$-$G027$^d$       & 66.0$^{+5.0}_{-4.8}$    & 1.280$^{+0.072}_{-0.075}$   & 2.67$^{+0.13}_{-0.14}$     & 6.412$^{+0.026}_{-0.022}$ & 53$^{+31}_{-36}$     &350$^{+120}_{-70}$ & 8.9$^{+2.0}_{-1.8}$      & 1.30$^{+0.17}_{-0.19}$   & 1.17(80)  \\  
 2MASX J12544196$-$3019224& 20.6$^{+7.0}_{-5.3}$      & 1.9(f)               & 0.45$^{+0.16}_{-0.12}$     & 6.40(f)                 &  10(f)              &130($<$410)      & 0.18($<$0.57)          & 0.63$^{+0.25}_{-0.21}$ & 743.7(1923)$^e$   \\ 
 NGC 4939                 & 19.3$^{+3.0 }_{-2.6}$   & 1.9(f)                  & 3.08$^{+0.52}_{-0.45}$     & 6.40(f)                 & 10(f)               &83($<$220)       & 0.76($<$2.01)          & 3.86$^{+0.66}_{-0.67}$    & 1.21(28) \\ 
 ESO 383$-$G18            & 19.46$^{+0.92}_{-0.86}$ & 1.545$^{+0.048}_{-0.046}$ & 2.597$^{+0.069}_{-0.067}$   &6.325$^{+0.054}_{-0.087}$ & 100$^{+76}_{-71}$     &130$^{+40}_{-53}$  & 1.95$^{+0.69}_{-0.77}$    & 2.68$^{+0.21}_{-0.23}$     & 1.30(134) \\ 
 ESO 103$-$G035           & 18.99$^{+0.48}_{-0.46}$ & 1.775$^{+0.040}_{-0.039}$ & 15.79$^{+0.24}_{-0.23}$     &6.476$^{+0.023}_{-0.022}$ & 3($<$47)             & 67$\pm19$       & 3.9$\pm1.1$             & 3.03$^{+0.34}_{-0.32}$    & 1.19(136)   \\
 IC 4995$^d$              & 59.6                  & 3.44                    & 11.1                     & 6.40                   & 0                    & 790             & 1.50                    & 2.80                    & 1.93(19)  \\
NGC 7070A                & 12.4$^{+0.7}_{-1.1}$    & 1.510$^{+0.070}_{-0.090}$    & 0.72$^{+0.22}_{-0.19}$     & 6.40(f)                & 10(f)                &38($<$77)       &0.17($<$0.34)             & 0.669$\pm0.090$         & 0.98(96)  \\
 NGC 7172                 & 7.810$^{+0.072}_{-0.070}$ & 1.656$^{+0.013}_{-0.015}$ & 15.56$\pm0.09$         &6.380$^{+0.021}_{-0.019}$  & 80$\pm25$            &71$^{+11}_{-9}$   & 5.20$^{+0.80}_{-0.70}$    & 3.62$\pm0.24$            & 1.30(173) \\  
 NGC 7319$^d$             & 66.1                 & 2.25                    & 6.14                   &6.39                   & 48                   &250              & 2.39                    & 2.81                    & 2.99(131)   
\enddata
\tablecomments{Photon index of the power law with only Galactic absorption was assumed to be the same value as power law absorbed by cold matter at the redshift of the source. (f) indicates fixed the parameter.}
\tablenotetext{a}{Normalization of the absorbed power law in units of $10^{-3}$photons cm$^{-2}$ s$^{-1}$ at 1 keV.}
\tablenotetext{b}{Normalization of the Gaussian line in units of $10^{-5}$ photons cm$^{-2}$ s$^{-1}$ in the line.}
\tablenotetext{c}{Normalization of the less absorbed power law in units of $10^{-5}$ photons cm$^{-2}$ s$^{-1}$ at 1 keV.}
\tablenotetext{d}{Errors were not calculated for the case of $\chi^2_{\rm \nu}$ exceeding 1.5.}
\tablenotetext{e}{$C$-statistic was used to fit the unbinned spectrum. $C$-statistics and number of bins (in parenthesis) are shown.}
\label{table:baseline}
\end{deluxetable*}
}

\subsection{Complex Models}
Spectra of 15 sources were not satisfactorily fitted with the baseline
model. They showed soft X-ray excess below $\sim$2 keV and/or
residuals between 2 and 4 keV. We added extra components modified by
Galactic absorption to the baseline model until good fits were
obtained.  The photon indices for NGC 4939 and IC 4995 were fixed at
1.9, since uncertainties of $\Gamma$ were large if $\Gamma$ was left
free.

First, we added an optically thin thermal plasma model ({\tt mekal}
model in XSPEC; Mewe et al. 1985; Kaastra. 1992; Liedahl et al. 1995)
to the baseline model. The abundance was fixed at 0.5 solar, where the
solar abundance table by Anders and Grevesse (1989) was assumed. If
the temperature of the plasma was not constrained, the value was fixed at
0.65 keV, which is typically observed in Seyfert 2 (e.g., Guainazzi et
al. 2005b). Spectra of six objects (NGC 1142, 3C 98, IC 2461, NGC
4138, NGC 4939, and NGC 7172) were fitted acceptably with this
model. In the spectrum of NGC 7172, excess emission at around 1.7 keV
was seen in this model fit ($\chi^2_{\rm \nu}$(dof) =
1.27(170)). We added a second Gaussian component in order to account
for this feature, and obtained an improved fit ($\chi^2_{\rm
  \nu}$(dof) = 1.18(168)). Soft excess seen in the spectrum of IC
4995 was modeled by adding a second {\tt mekal} component
($\chi^2_{\rm \nu}$(dof) = 1.14(15)). The best-fit temperatures of
the two {\tt mekal} components are $kT\approx$ 0.08 and $\approx$ 0.5
keV, respectively.

Next, we added a Compton reflection model ({\tt pexrav} in XSPEC;
Magdziarz \& Zdziarski 1995) to the baseline model.  We fixed the
inclination angle of the reflector at 60$^{\circ}$ (0$^\circ$
corresponds to face-on), the high-energy cutoff of the incident power
law at 300 keV, and assumed solar abundances (Anders \& Grevesse
1989). The {\tt pexrav} model was used in such a way that it produces
reflected emission only (\textit{rel}$_{\rm refl}$ parameter was set
to $-$1 in XSPEC) since the direct component was modeled by an
absorbed power law. The normalization and $\Gamma$ of {\tt pexrav}
were assumed to be the same as those of the absorbed power-law
component. This model was applied to the remaining eight objects. If
the absorption column for the reflection component was left free, the
spectrum of ESO 506$-$G027 was reproduced by this model. The spectrum
of other objects was not explained by adding this model component.

In order to represent the spectra of the remaining seven sources, both
{\tt mekal} and {\tt pexrav} were added to the baseline model. The
spectra of 3C 33, and ESO 383$-$G18 were explained by this model. If
the absorption column for the reflection component was left free, the
spectra of ESO 103$-$G035 and NGC 7319 were reproduced. If two {\tt
  mekal} components ($kT\approx$ 0.2 and $\approx$ 0.8 keV) and
absorbed pexrav were introduced, the spectrum of NGC 4388 was
explained.

Mrk 348 and NGC 4507 show more complex X-ray spectral shapes that were
not reproduced by the models explained above. In the fit of the
spectrum of Mrk 348, we added two {\tt mekal} components and the third
absorbed power law to the baseline model. The photon indices of all
the power-law components were assumed to be the same. Moreover, two
Gaussians with negative normalization were also added to represent two
absorption-line like features seen at 6.6 keV and 6.9 keV. This model
provided an acceptable fit to the spectrum of Mrk 348.  In order to
represent the spectrum of NGC 4507, we introduced the following model
components.  We used nine Gaussians instead of {\tt mekal} to model
the soft part of the spectrum. A Gaussian to express the Compton
shoulder (CS; see Matt 2002) was also added at 6.32 keV with a fixed
width of $\sigma$ = 40 eV as in Matt et al. (2004).  An absorbed {\tt
  pexrav} component was also added, where the absorption column
density for {\tt pexrav} was assumed to be independent of that for the
absorbed power law component.  The combination of the baseline model
and these additional components reproduced the spectrum.

The results of the fits are summarized in Tables 3, 4, and 5. Table
\ref{table:model} shows the best-fit models for our sample. The photon
indices were distributed between $\sim$1.5 and 2.0.  This is
consistent with a range of photon indices observed in Seyfert 2s
(e.g., Smith \& Done 1996). The obtained $N_{\rm H}$ were in the range
of $\sim$10$^{23-24}$ cm$^{-2}$. The spectra along with the best-fit
model are shown in Figure 2.

\subsection{Fluxes and Luminosities}
The X-ray fluxes calculated using the best-fit model are summarized in
Table \ref{table:flux}. Columns 2 and 3 are observed fluxes in the
0.5$-$2 keV and 2$-$10 keV bands, respectively. Columns 4 and 5
are observed fluxes for the power-law components in the 0.5$-$2 keV
and 2$-$10 keV bands, respectively. Columns 6 and 7 are absorption
corrected fluxes for the power-law components in the 0.5$-$2 keV and
2$-$10 keV bands, respectively. Columns 8 and 9 are observed and
absorption corrected fluxes in the 0.5$-$2 keV band, respectively, for
the power law with only Galactic absorption. Columns 10 and 11 are
observed and absorption corrected fluxes, respectively, for the {\tt
  mekal} component in the 0.5$-$2 keV band. We calculated intrinsic
luminosities of the absorbed power-law component in the 2$-$10 keV
band as shown in Table 8, Column 2. Most of the objects have
luminosities in the range of Seyferts (10$^{41-44}$ erg s$^{-1}$). The
intrinsic 0.5$-$2 keV luminosities for all the components except for
the heavily absorbed power law were also calculated and tabulated in
Table 8, Column 3.

\begin{figure*}
\centering
\includegraphics[width=7.5cm,height=8.5cm,clip,angle=270]{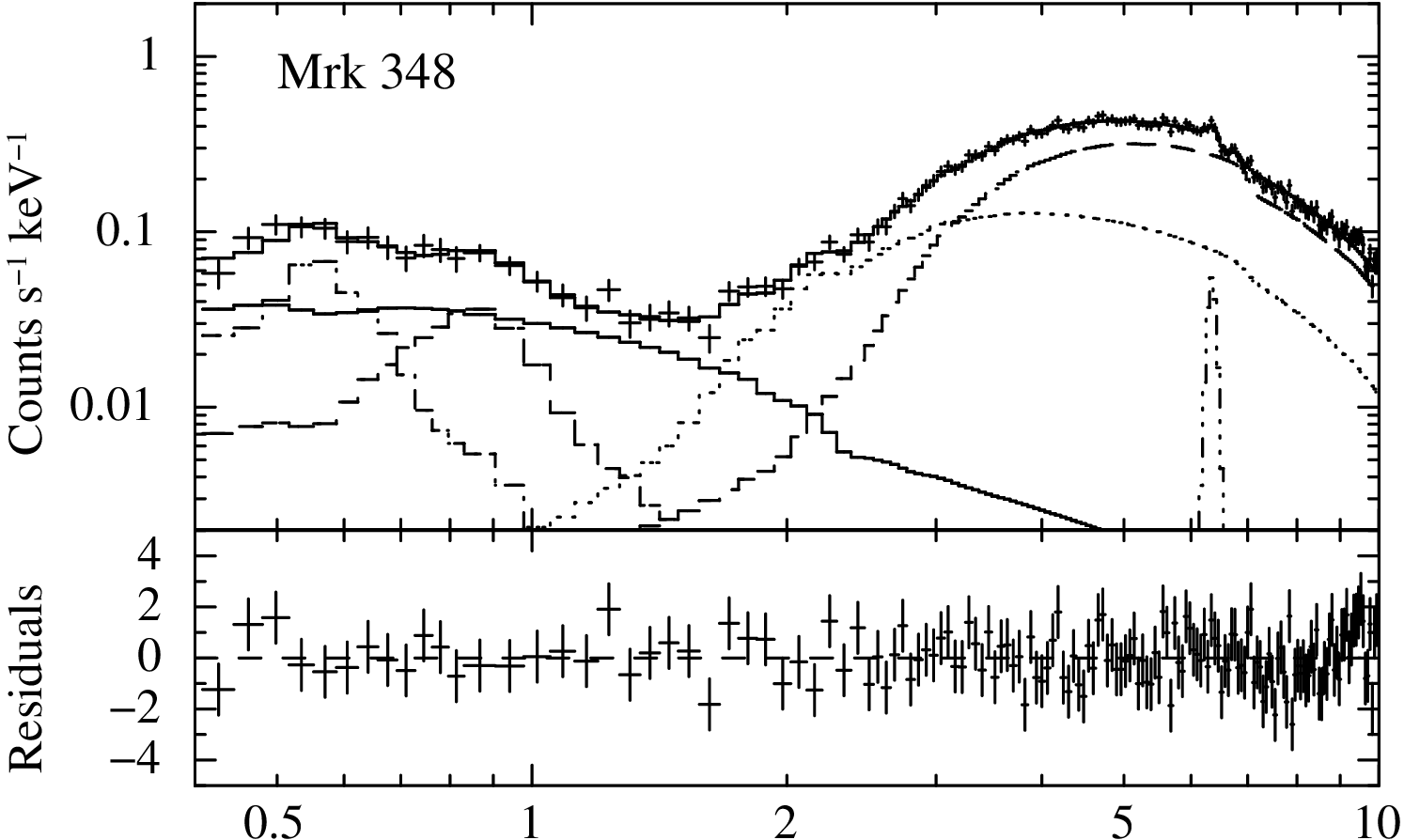}
\includegraphics[width=7.5cm,height=8.5cm,clip,angle=270]{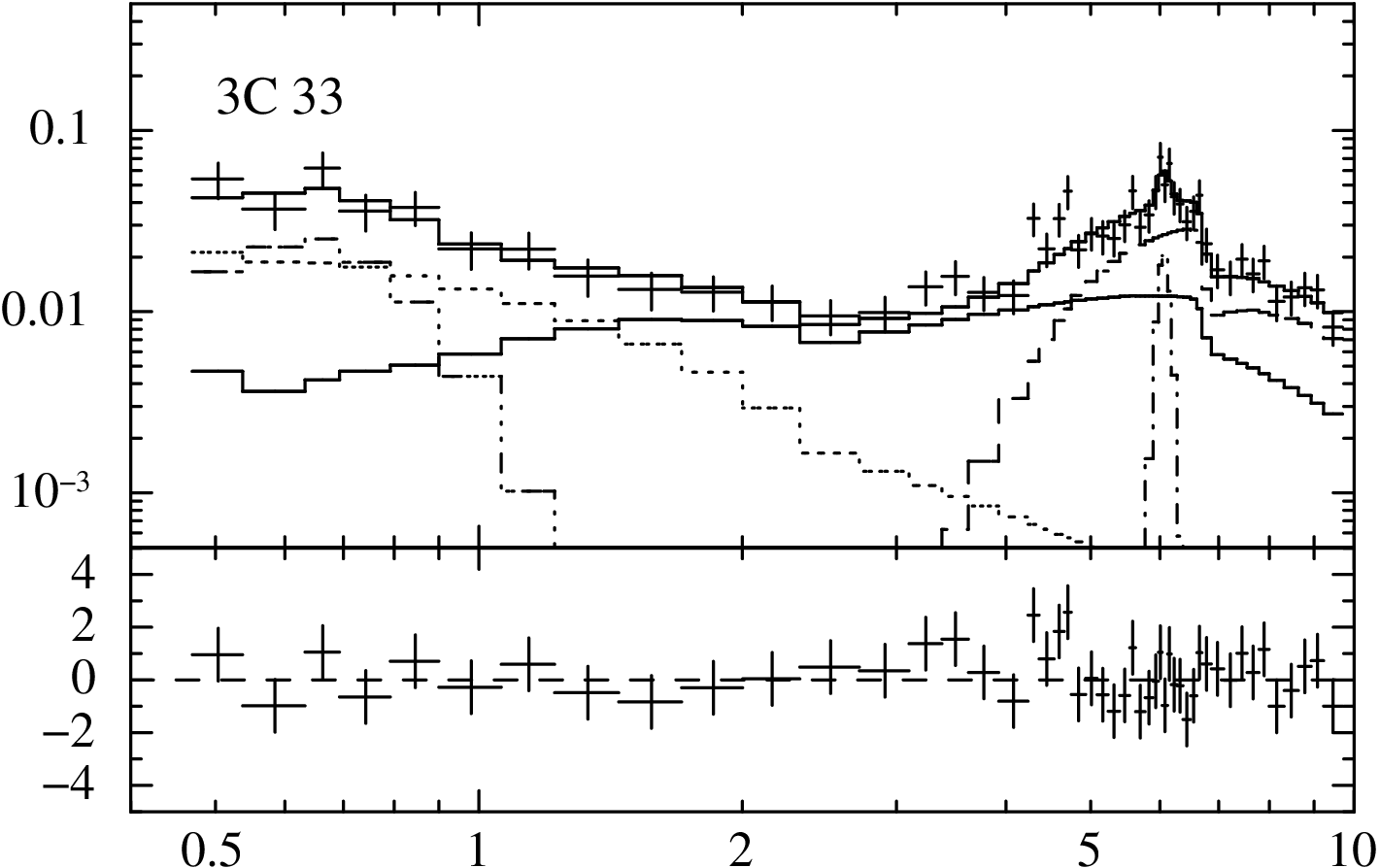}
\includegraphics[width=7.5cm,height=8.5cm,clip,angle=270]{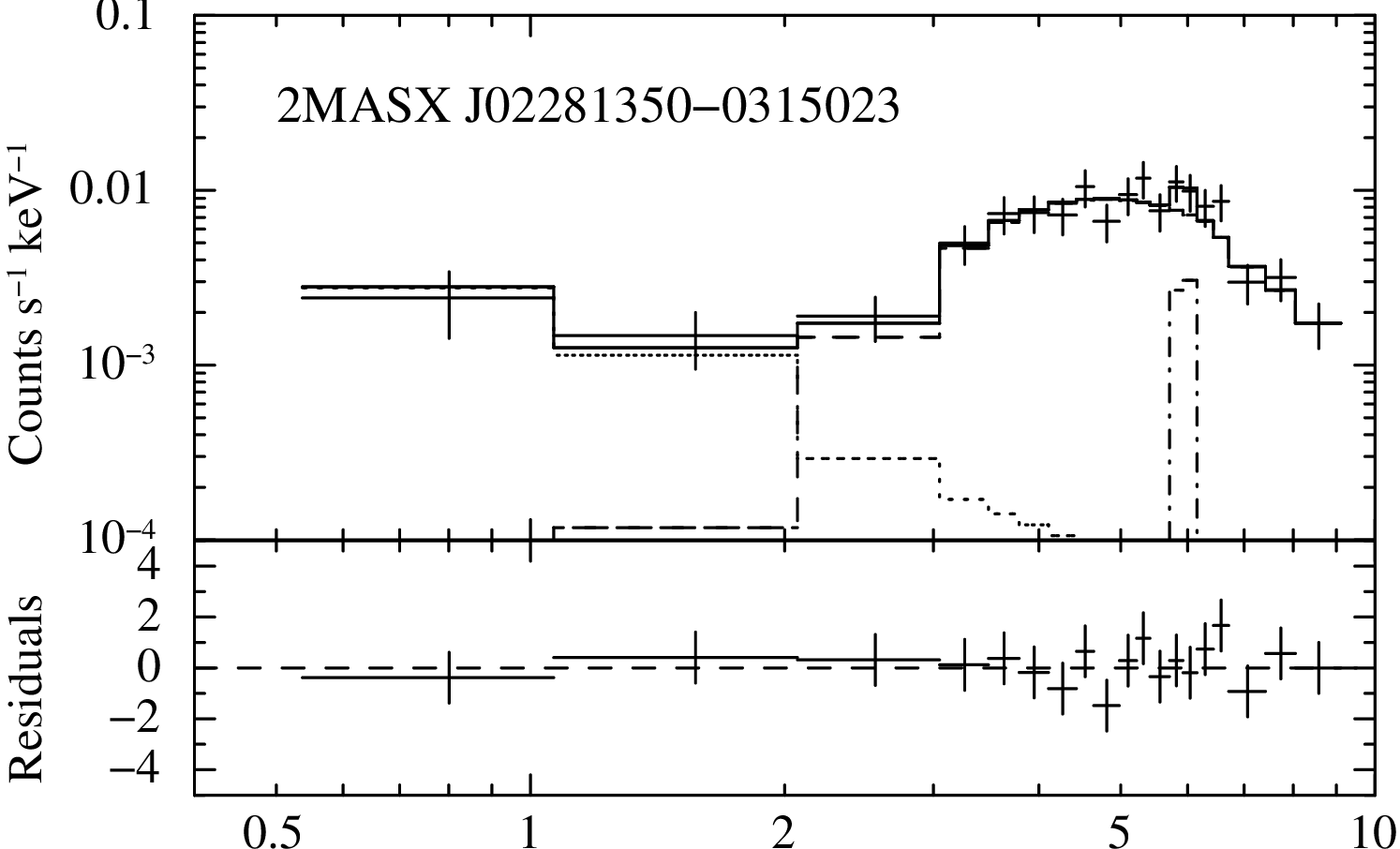}
\includegraphics[width=7.5cm,height=8.5cm,clip,angle=270]{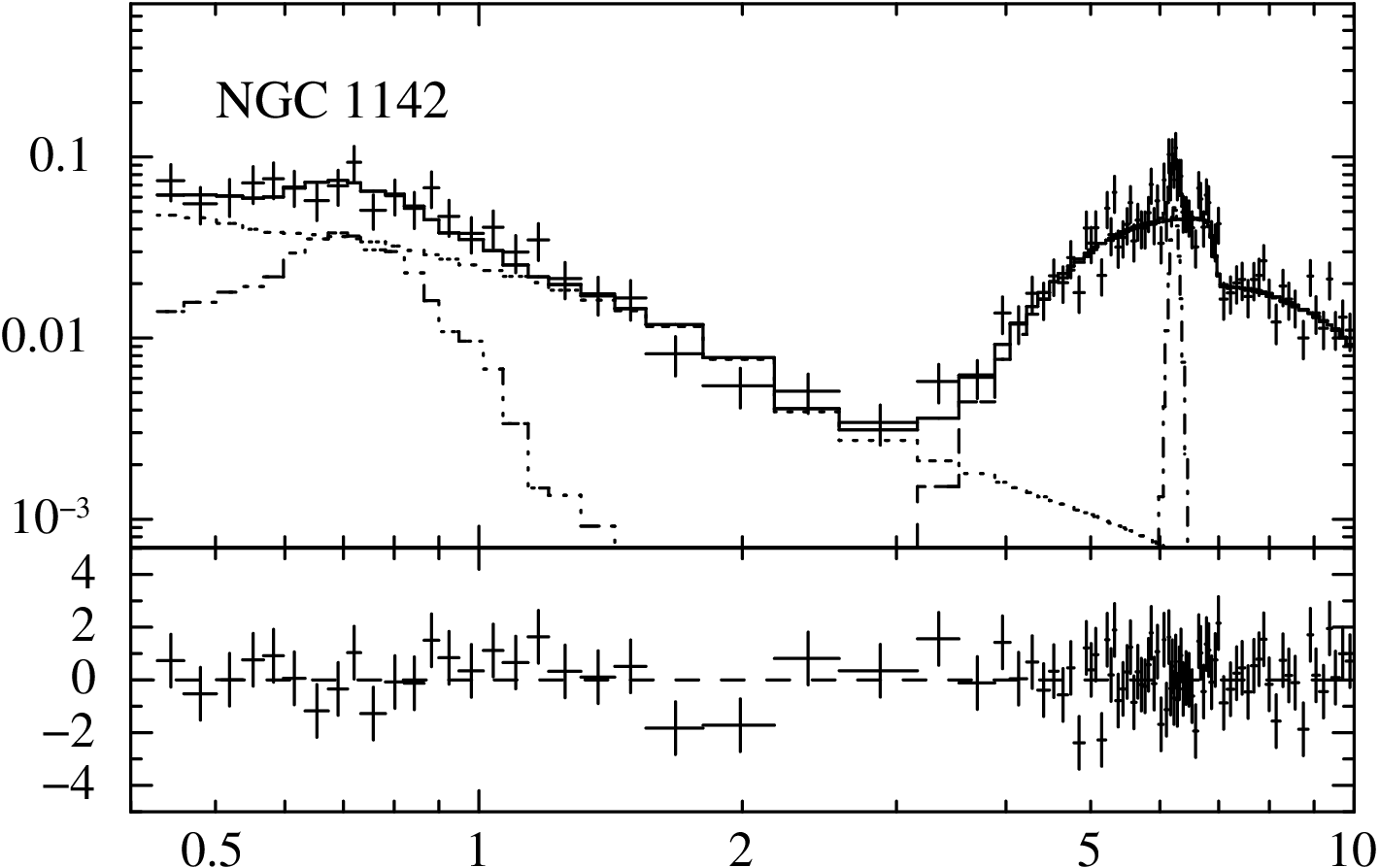}
\includegraphics[width=7.5cm,height=8.5cm,clip,angle=270]{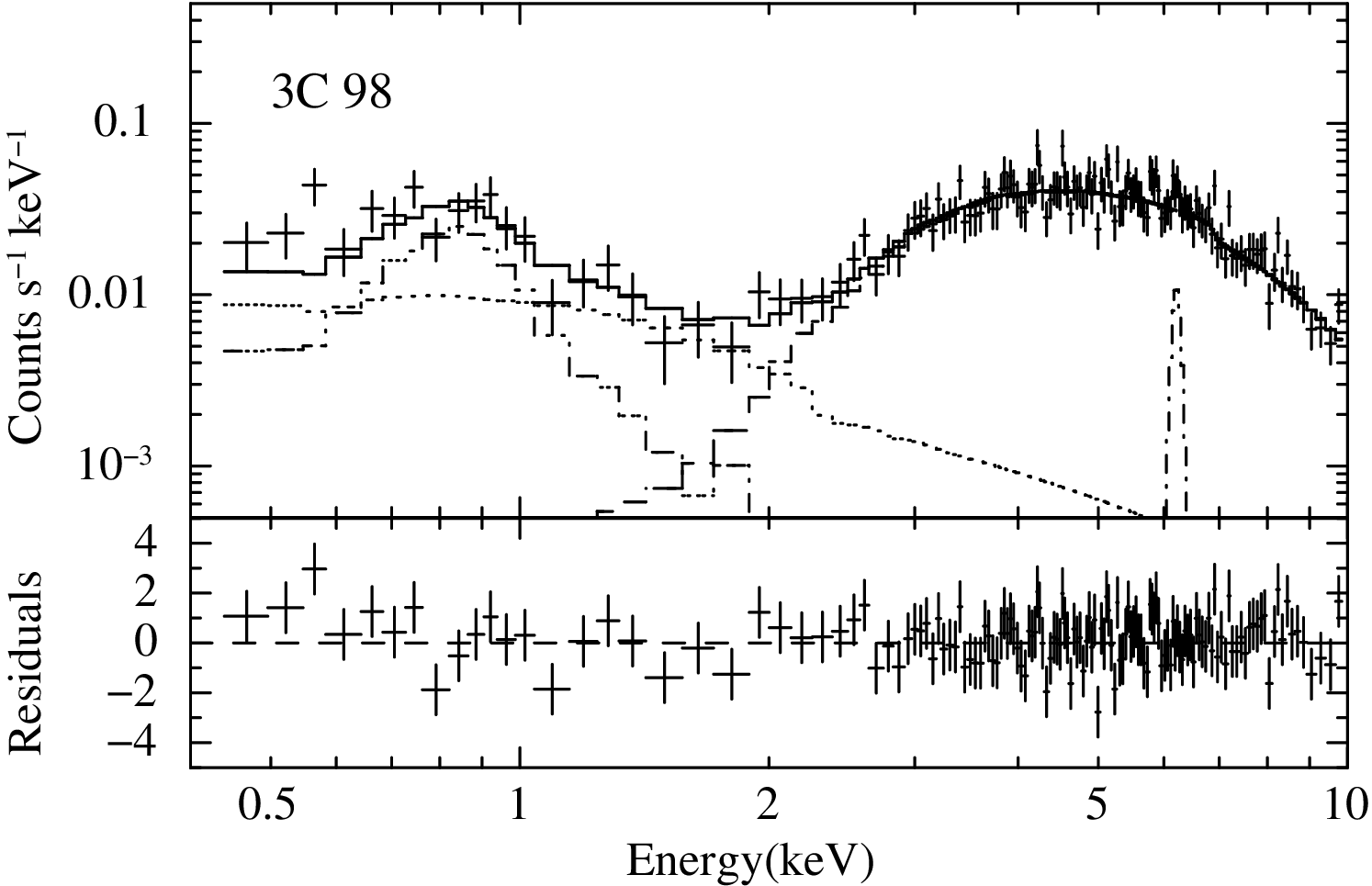}
\includegraphics[width=7.5cm,height=8.5cm,clip,angle=270]{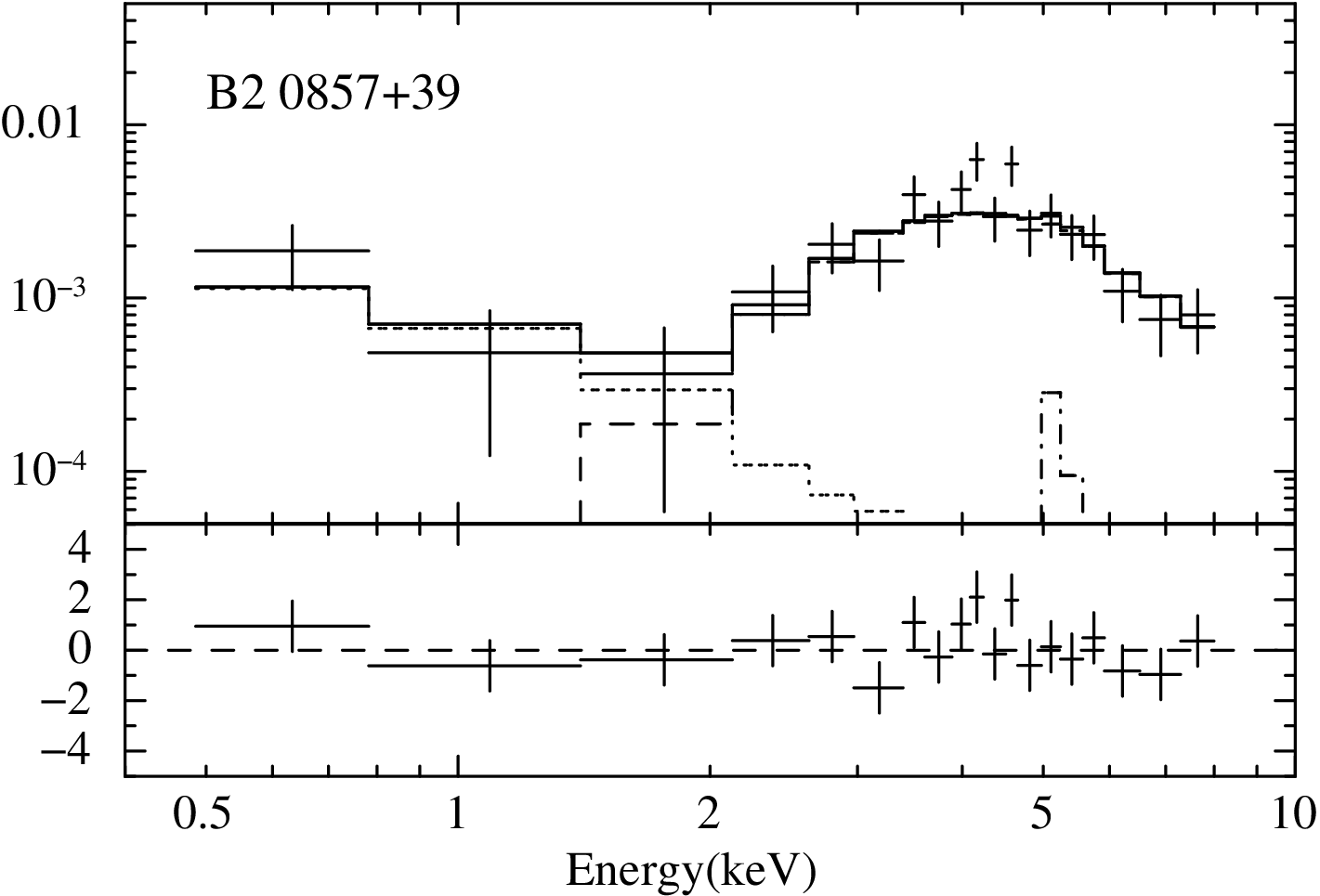}
\caption{X-ray spectra ({\it upper panels}) and residuals in units of
  $\sigma$ (or data/model ratio; {\it lower panels}). Model
  components are shown with dashed, dotted, dot-dashed, and
  dot-dot-dot-dashed lines. Although an unbinned spectrum was used
  in the fit of 2MASX J12544196$-$3019224, a binned spectrum is
  shown for the presentation purpose.}
\end{figure*}
\begin{figure*}
\centering
\figurenum{2}
\includegraphics[width=7.5cm,height=8.5cm,clip,angle=270]{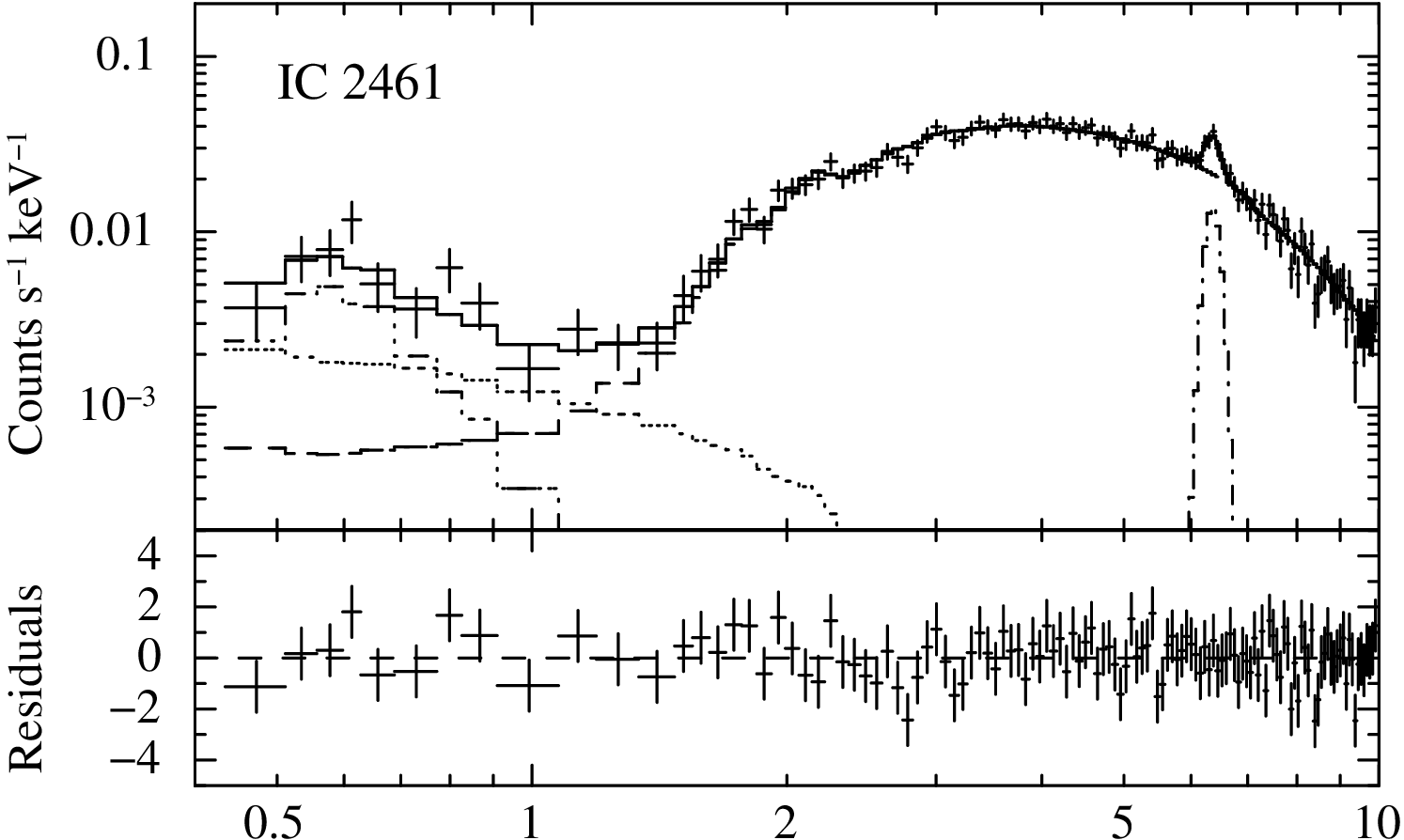}
\includegraphics[width=7.5cm,height=8.5cm,clip,angle=270]{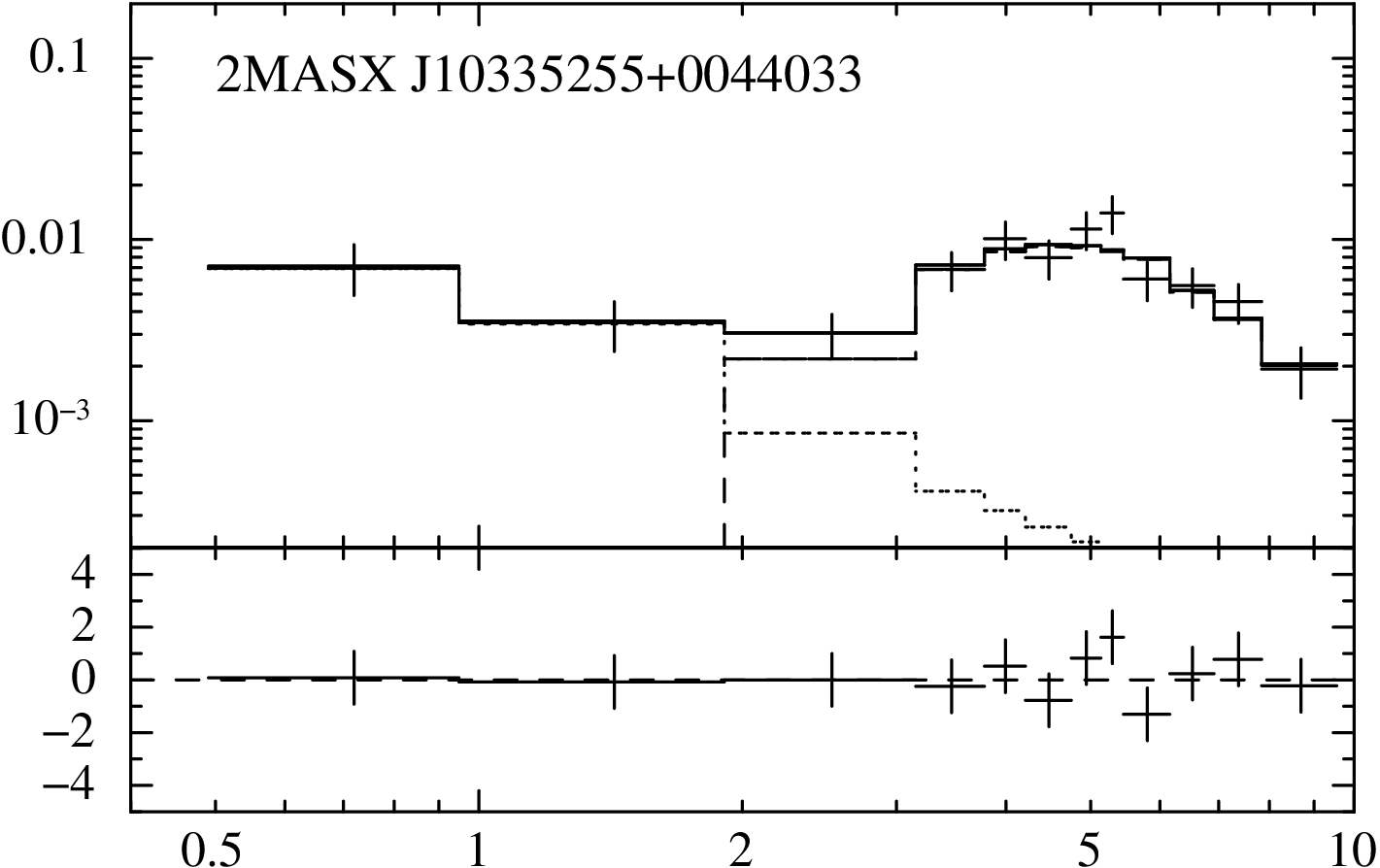}
\includegraphics[width=7.5cm,height=8.5cm,clip,angle=270]{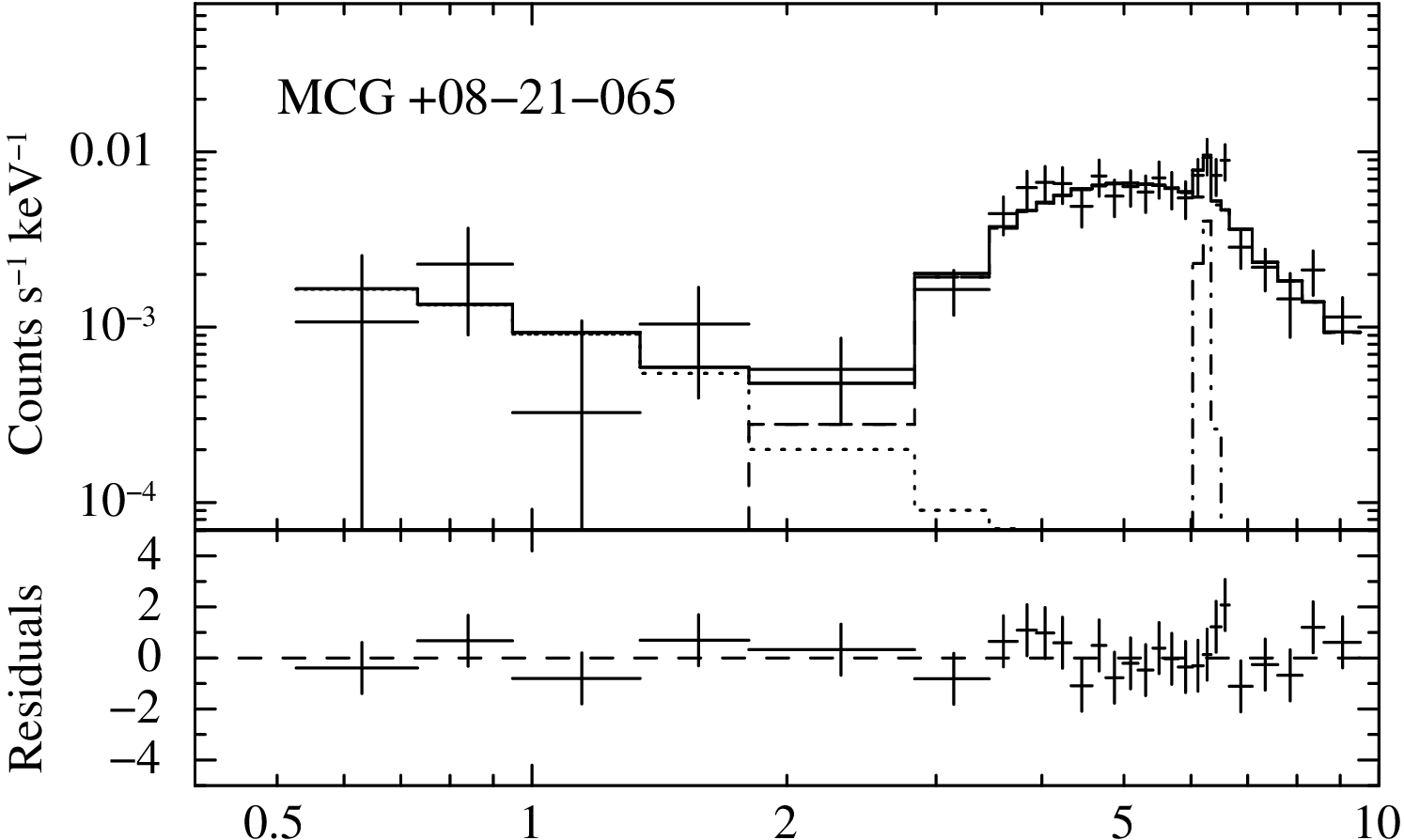}
\includegraphics[width=7.5cm,height=8.5cm,clip,angle=270]{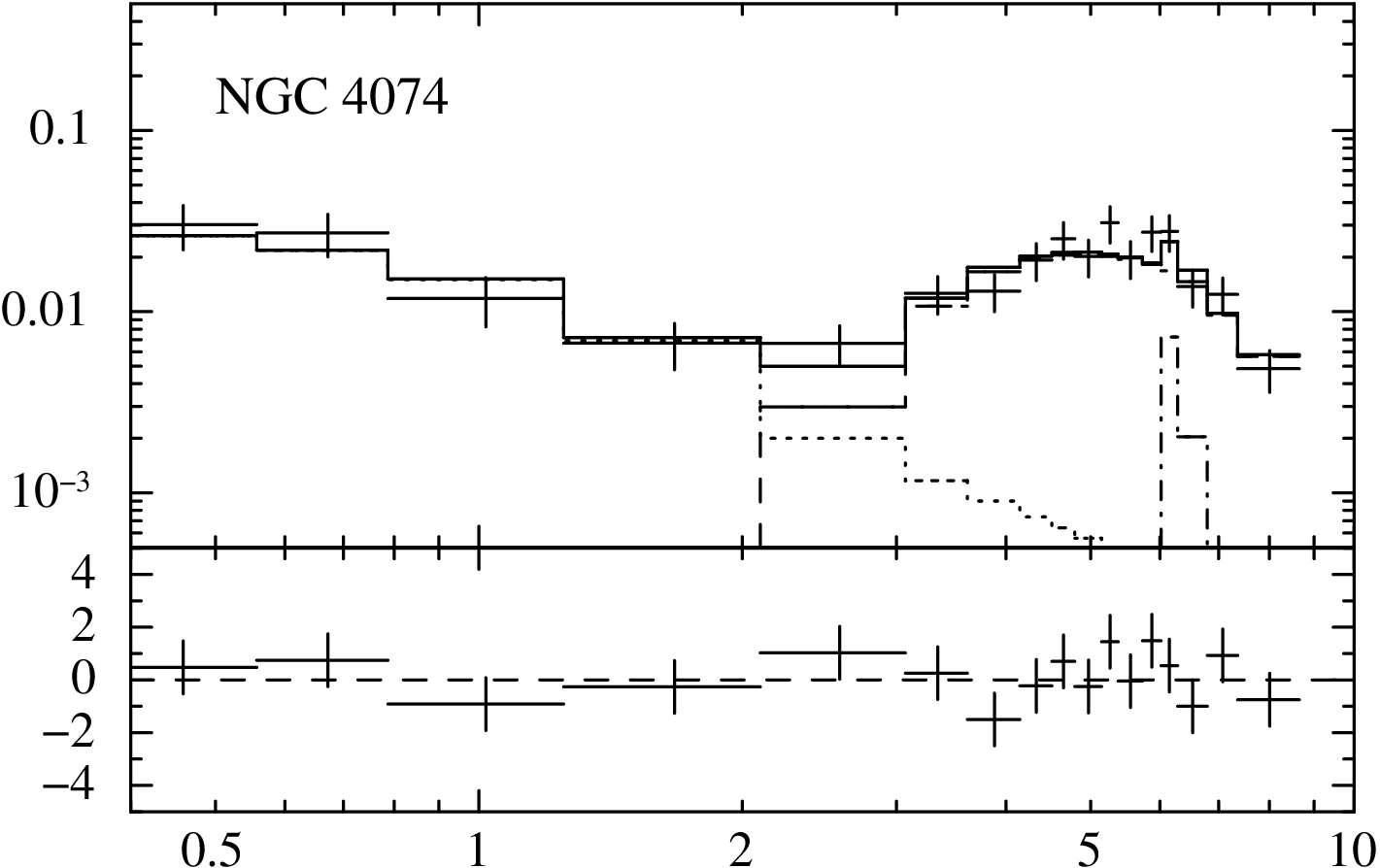}
\includegraphics[width=7.5cm,height=8.5cm,clip,angle=270]{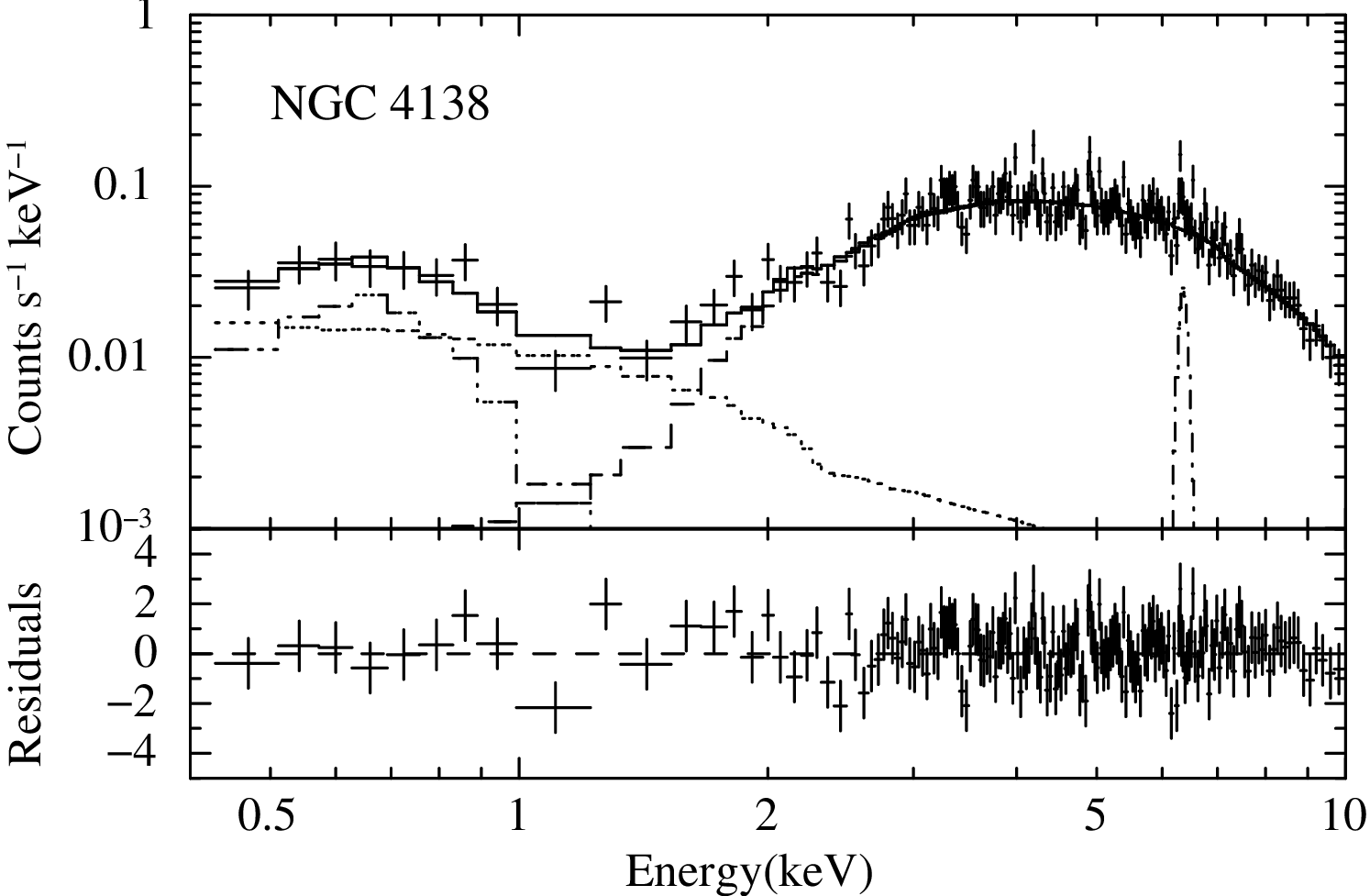}
\includegraphics[width=7.5cm,height=8.5cm,clip,angle=270]{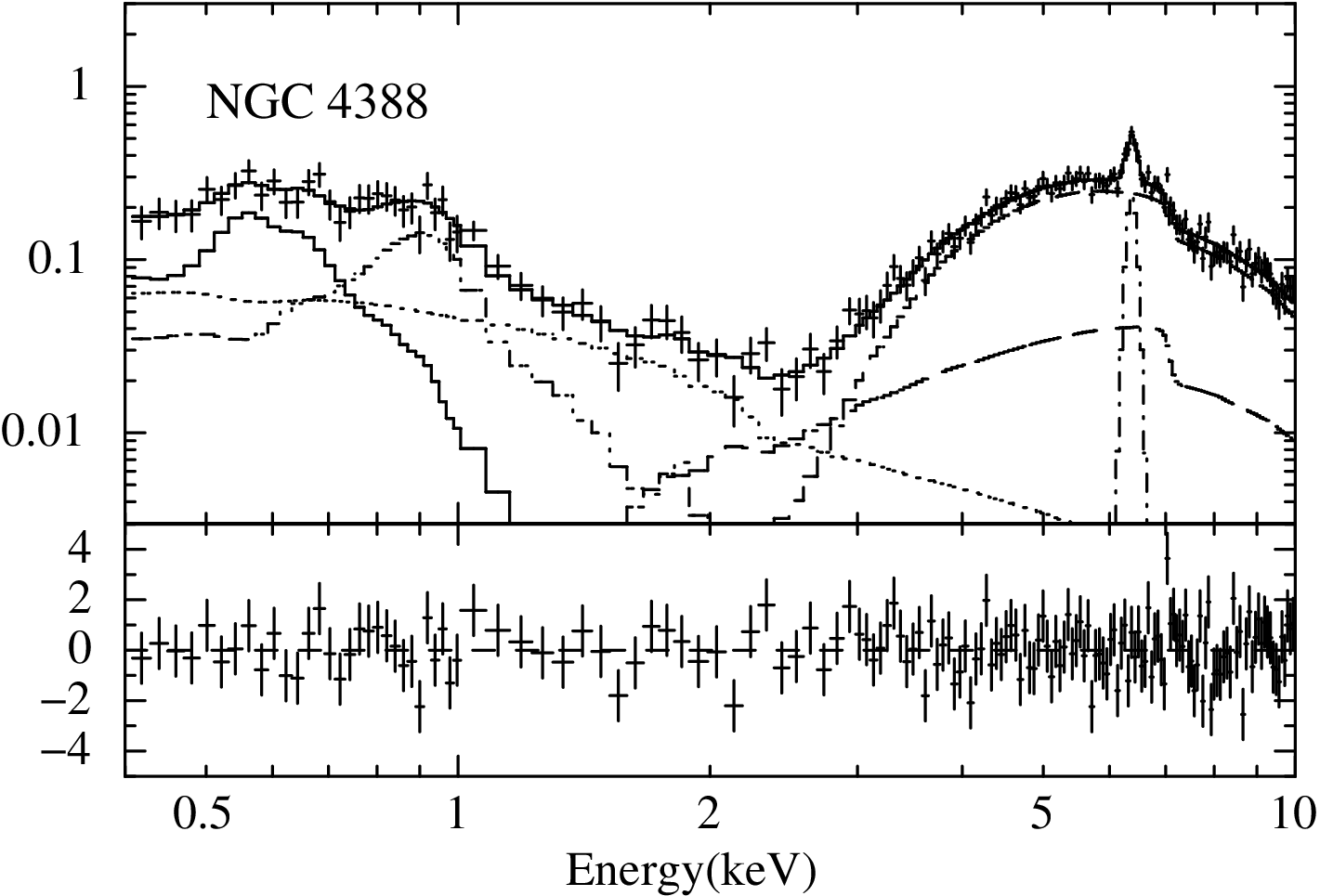}
\caption{Continued}
\end{figure*}
\begin{figure*}
\centering
\figurenum{2}
\includegraphics[width=7.5cm,height=8.5cm,clip,angle=270]{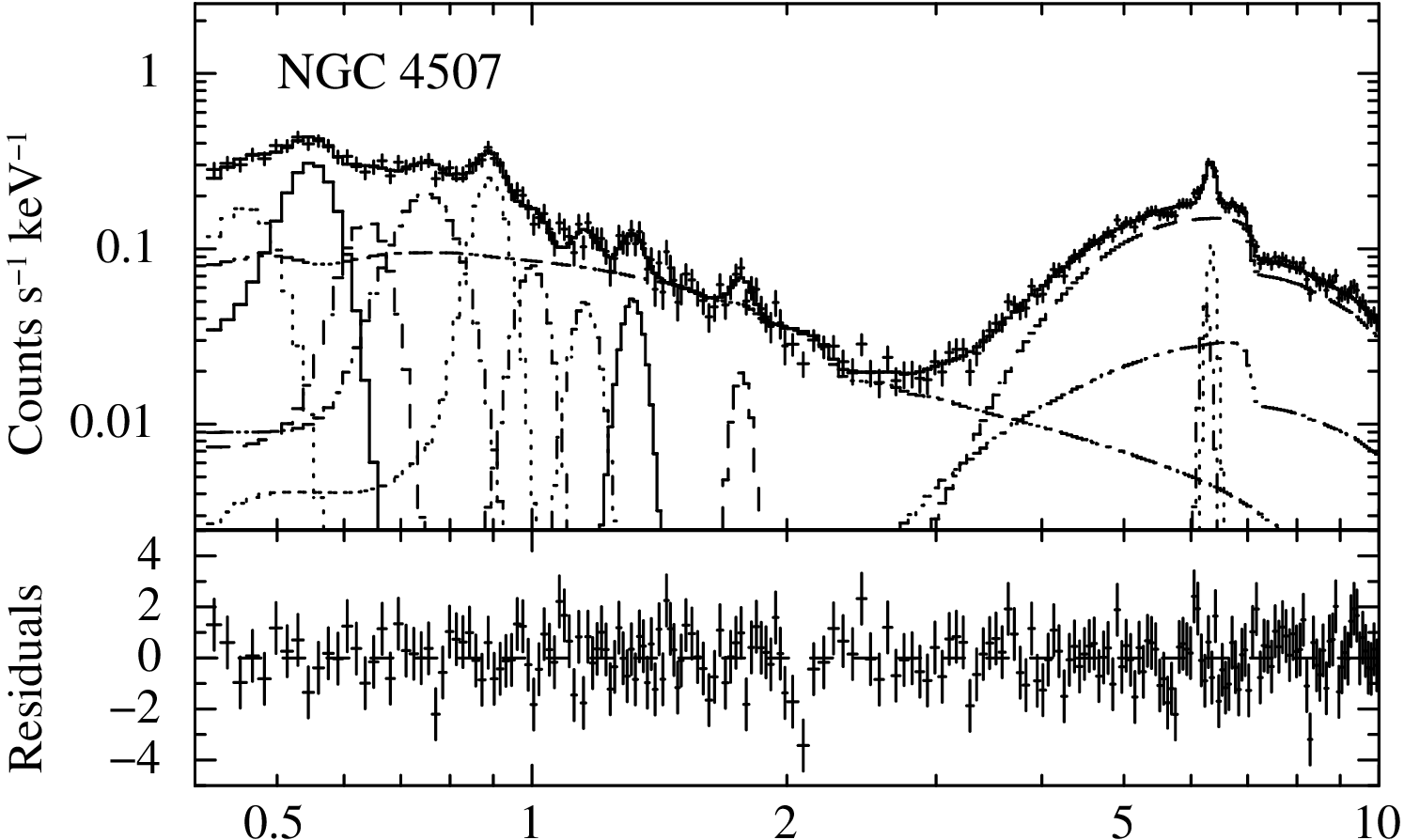}
\includegraphics[width=7.5cm,height=8.5cm,clip,angle=270]{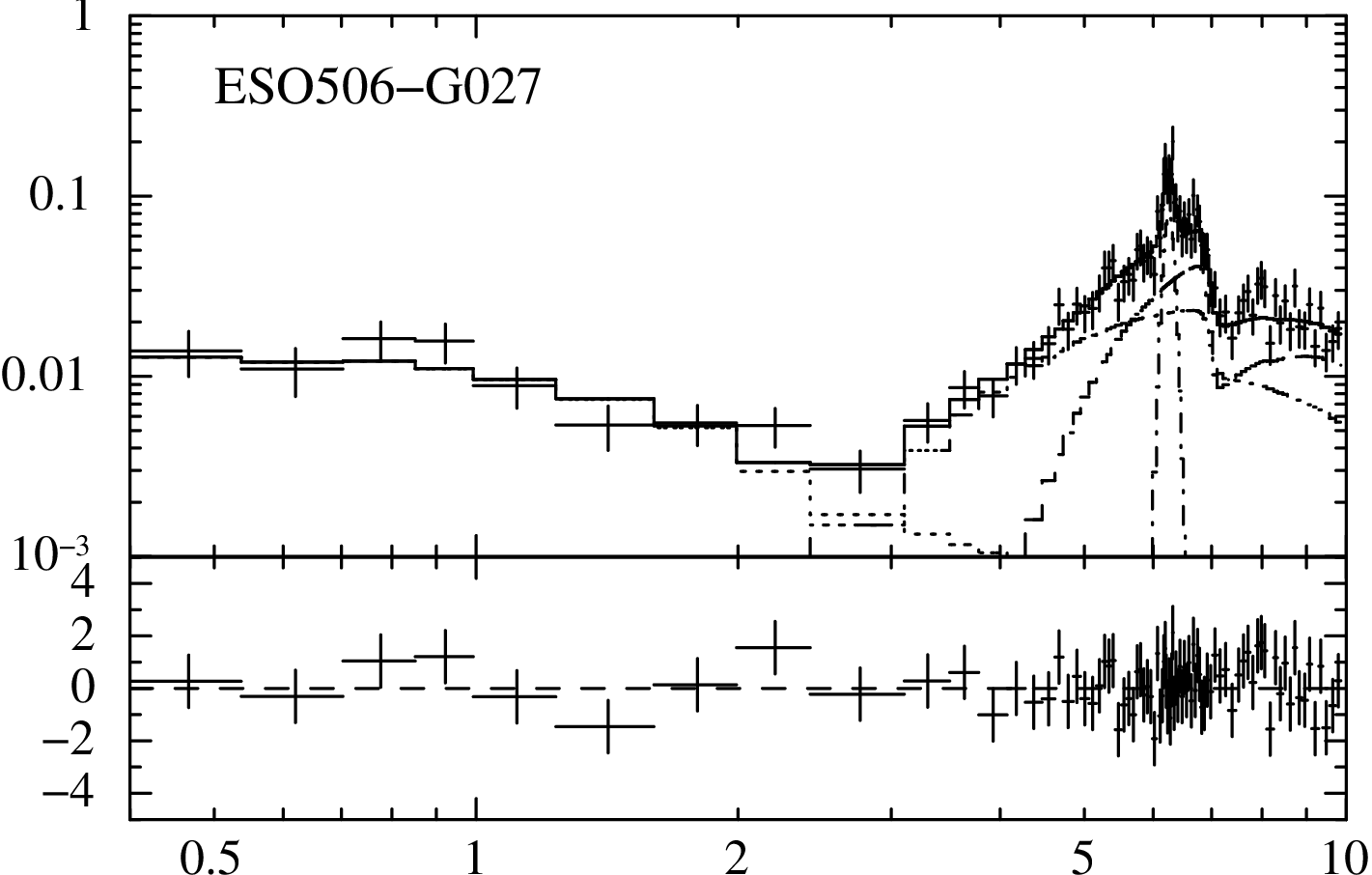}
\includegraphics[width=7.5cm,height=8.5cm,clip,angle=270]{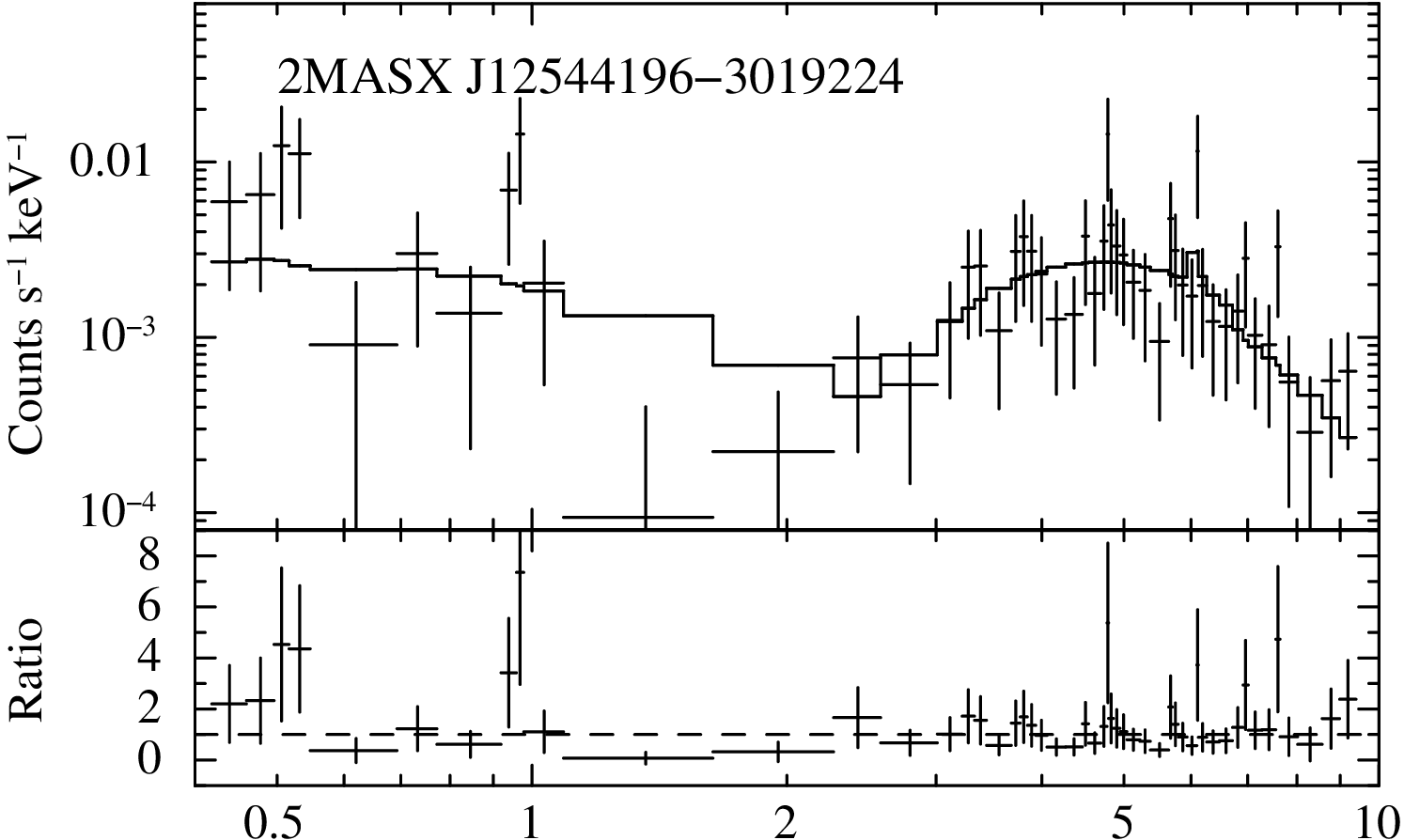}
\includegraphics[width=7.5cm,height=8.5cm,clip,angle=270]{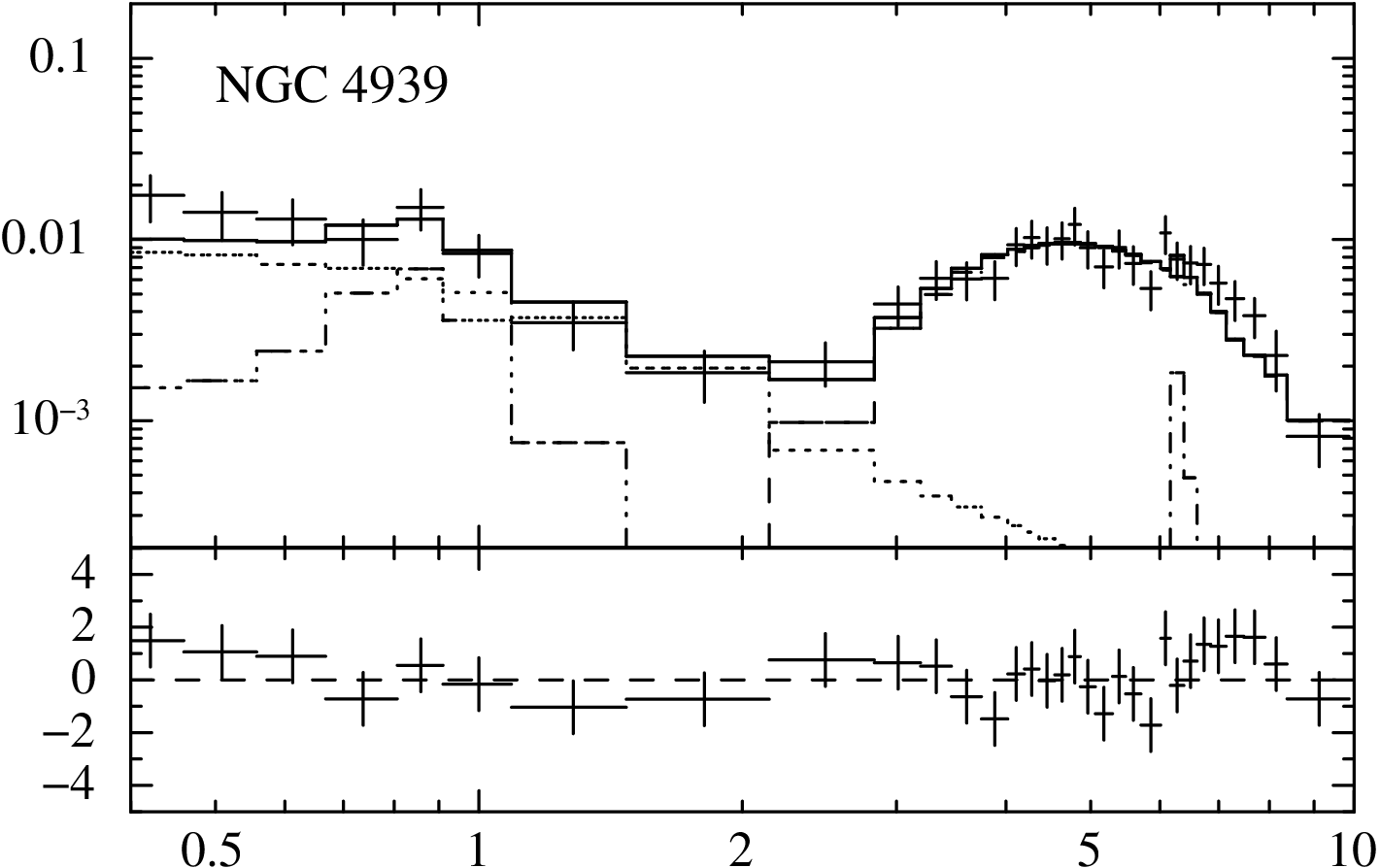}
\includegraphics[width=7.5cm,height=8.5cm,clip,angle=270]{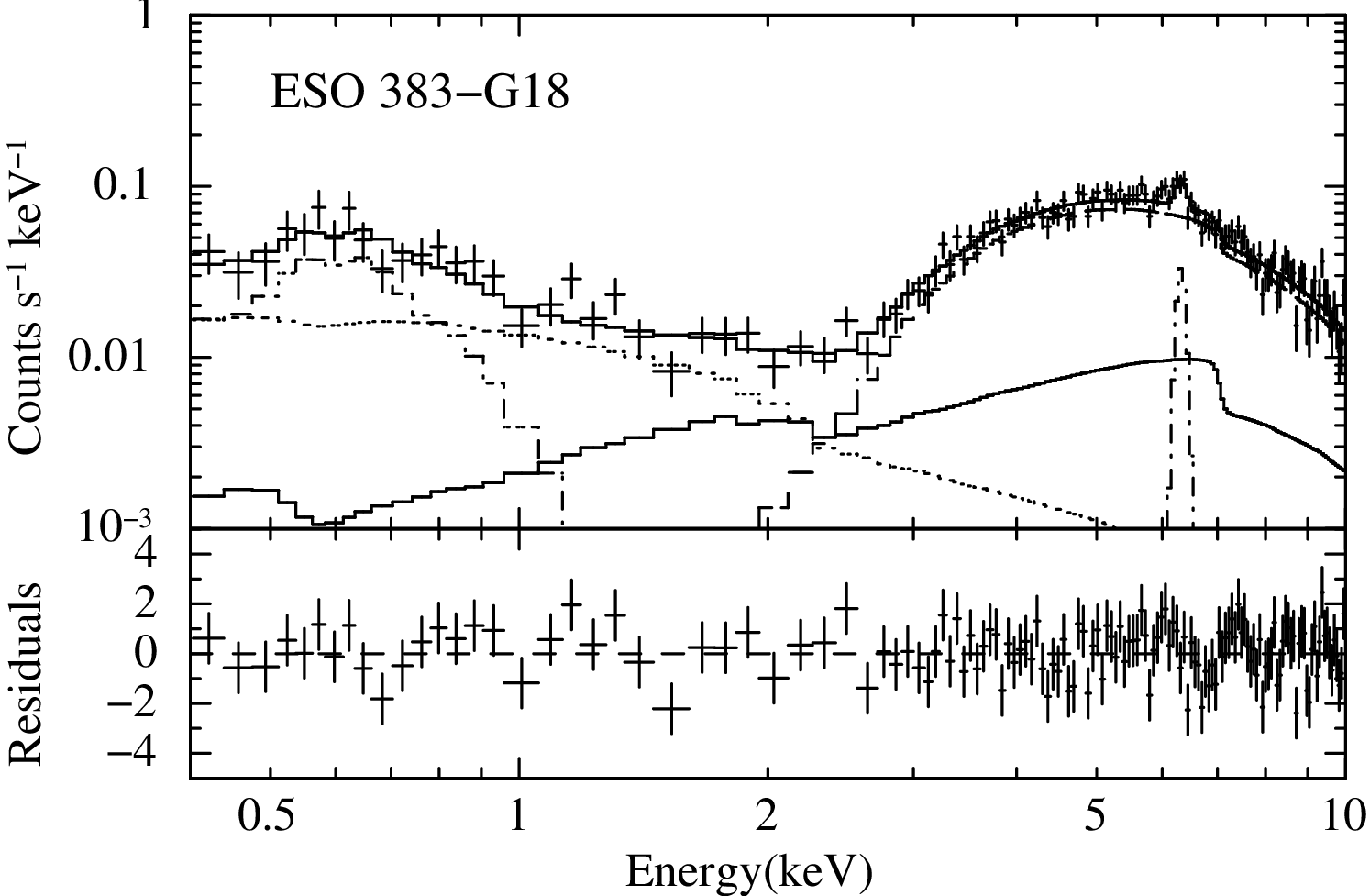}
\includegraphics[width=7.5cm,height=8.5cm,clip,angle=270]{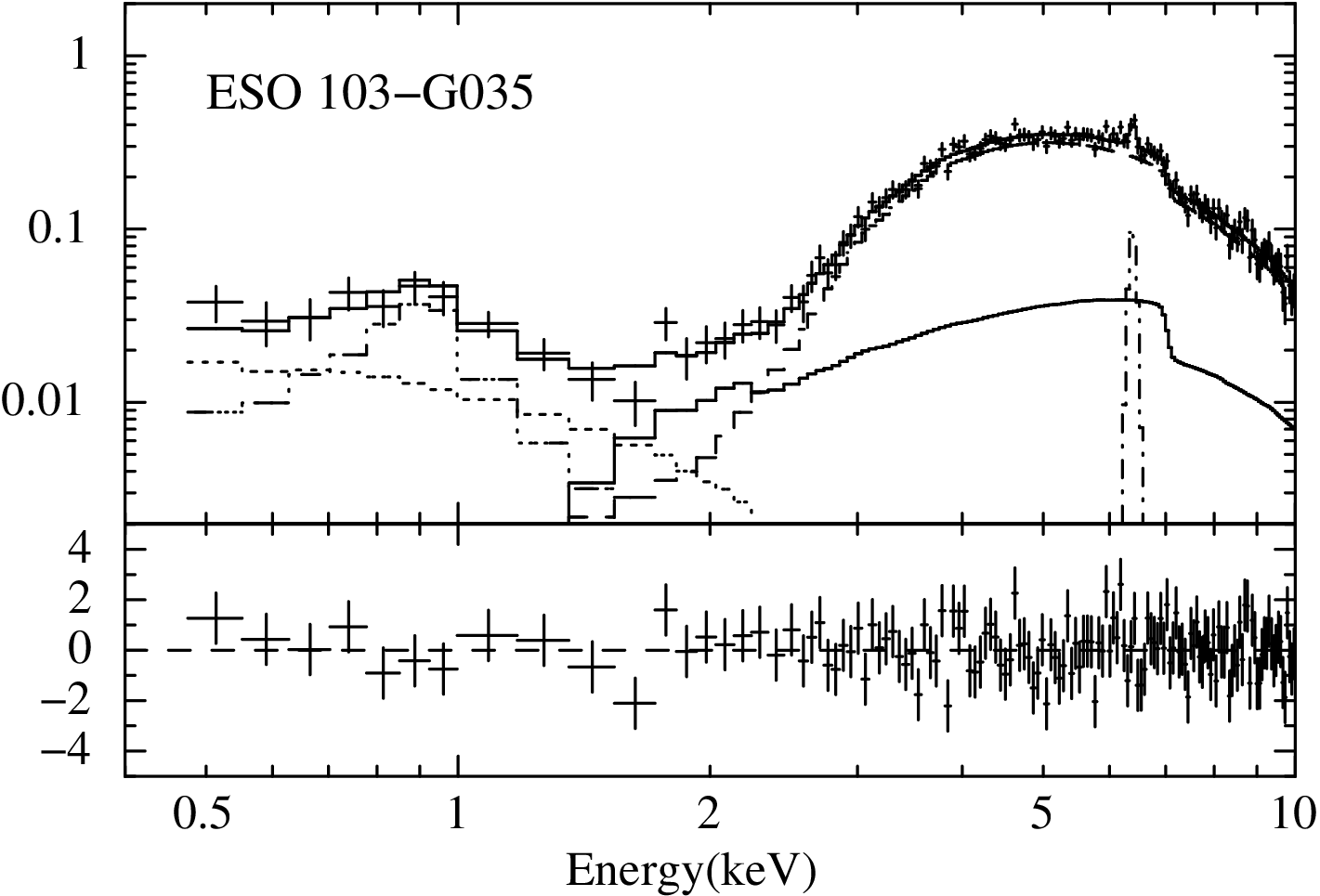}
\caption{Continued}
\end{figure*}
\begin{figure*}
\centering
\figurenum{2}
\includegraphics[width=7.5cm,height=8.5cm,clip,angle=270]{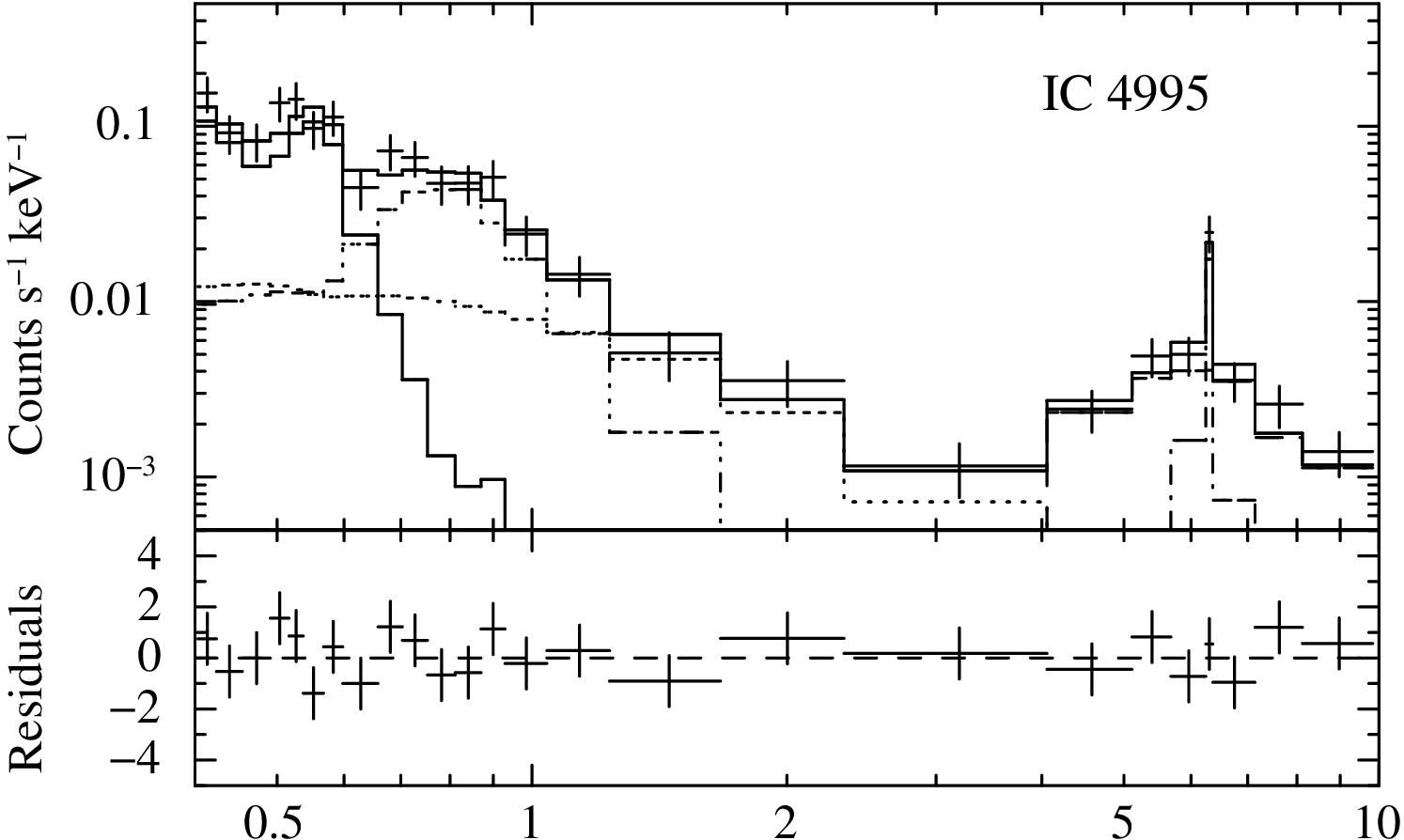}
\includegraphics[width=7.5cm,height=8.5cm,clip,angle=270]{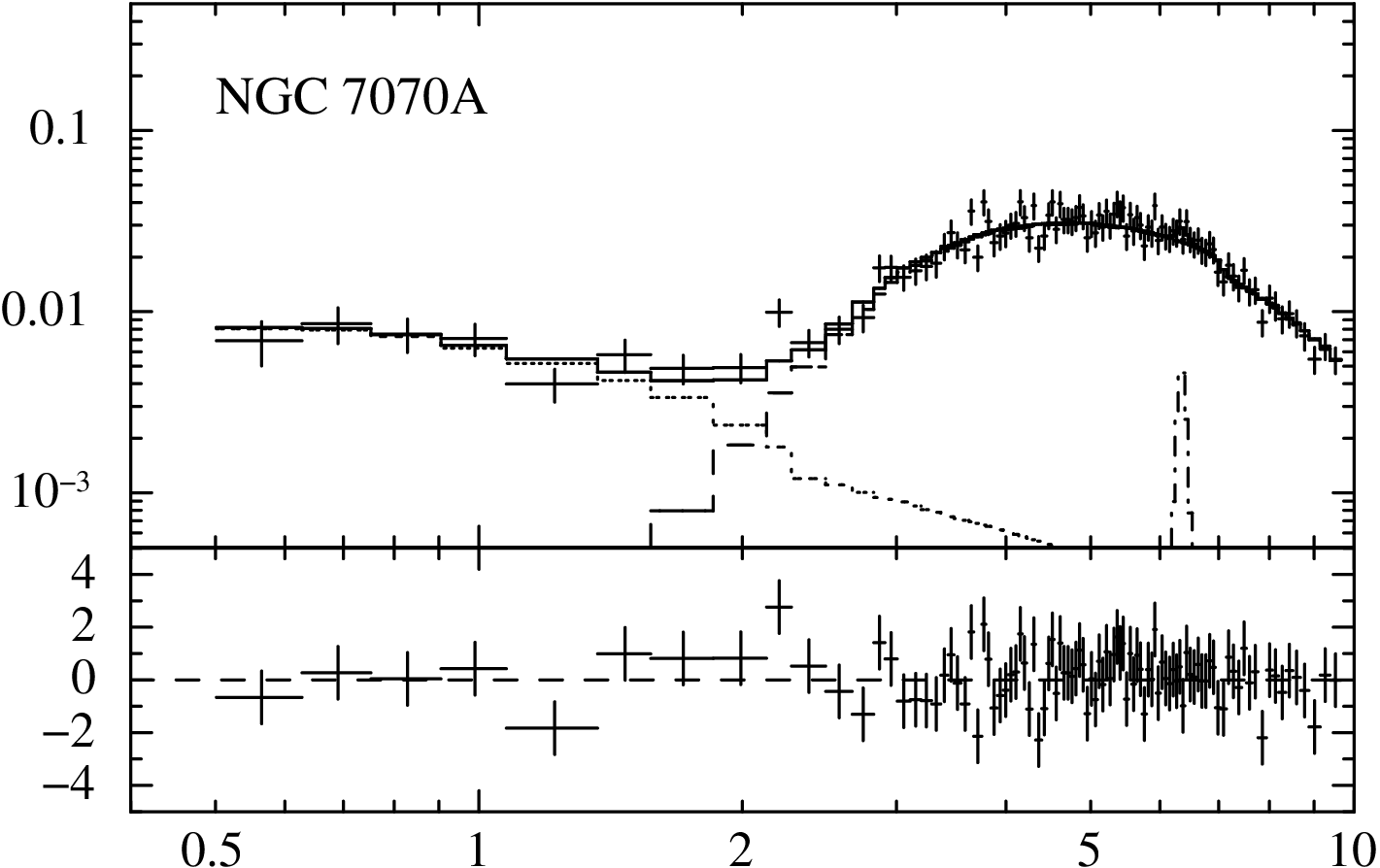}
\includegraphics[width=7.5cm,height=8.5cm,clip,angle=270]{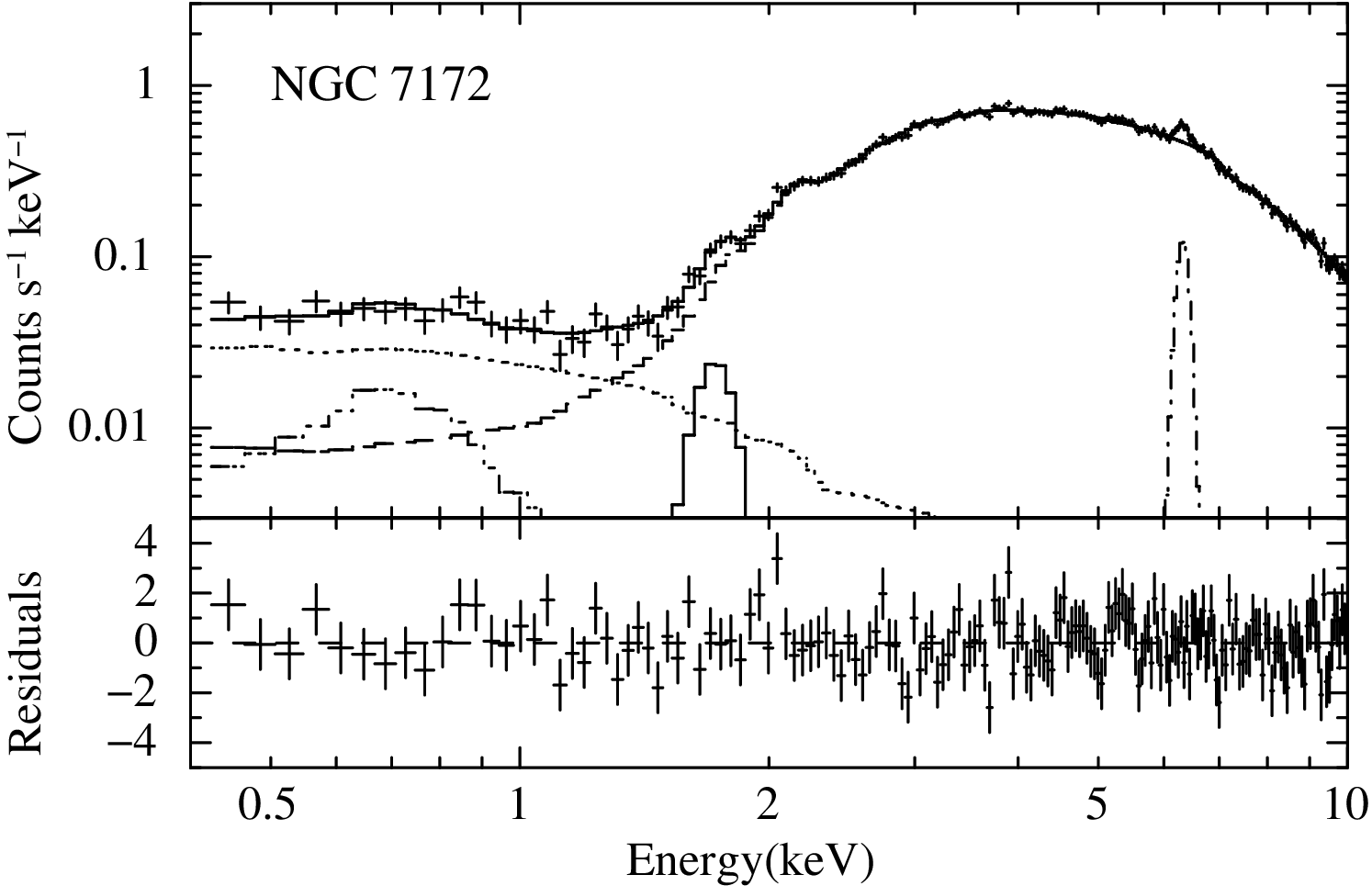}
\includegraphics[width=7.5cm,height=8.5cm,clip,angle=270]{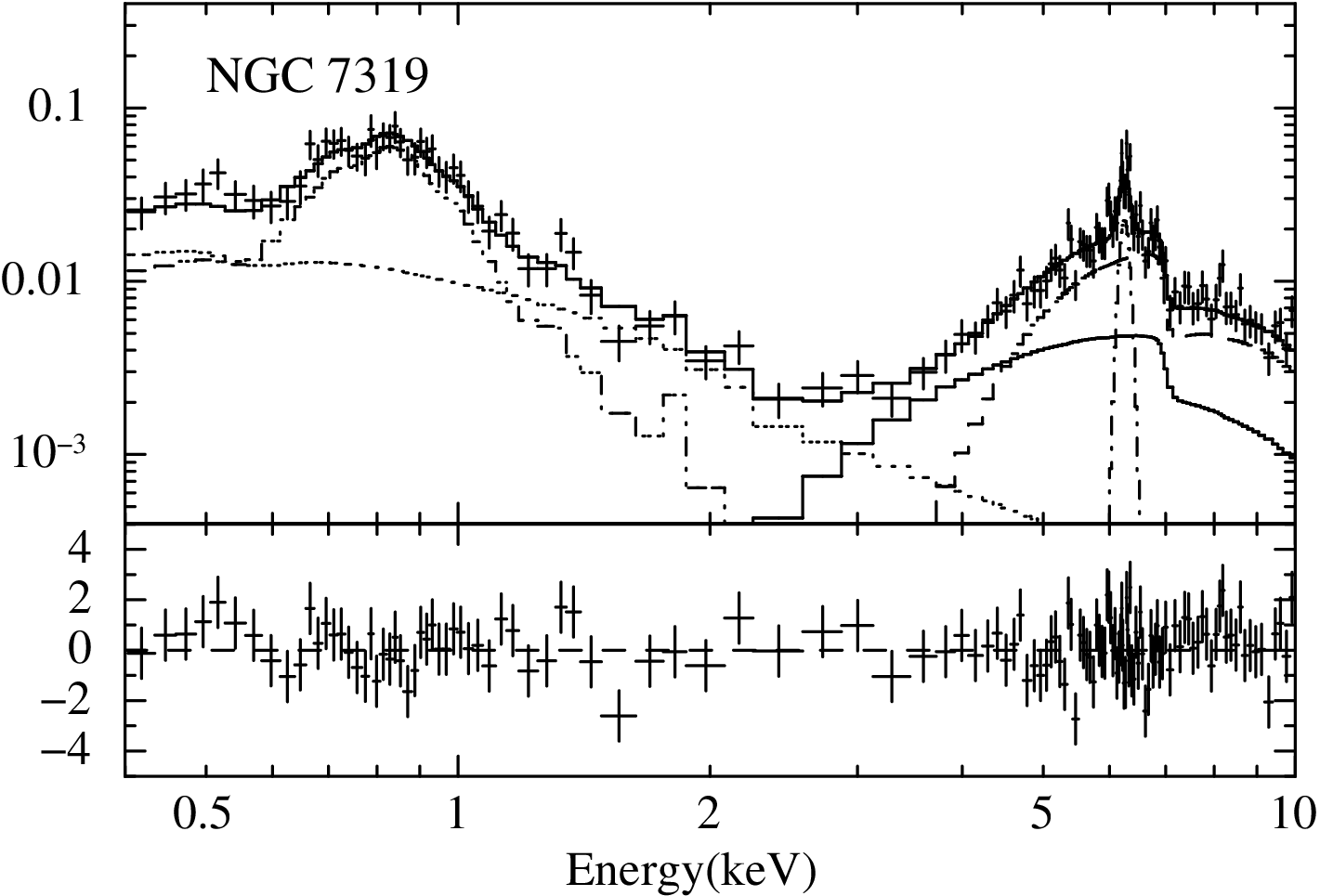}
\caption{Continued}
\end{figure*}

{\tabletypesize{\scriptsize}
\begin{deluxetable*}{lccccccccc}
\tablecolumns{10}
\tablecaption{Spectral Parameters for Absorption, Power Law, and Gaussian in the Complex Models}
\setlength{\tabcolsep}{-3pt}
\tablewidth{0pt}
\tablehead{
\colhead{}\\    
 \colhead{Name}              & \colhead{$N_{\rm H}$}      & 
 \colhead{$\Gamma$}          & \colhead{$A_{\rm int}$$^a$}          & 
 \colhead{$E_{\rm line}$}     & \colhead{$\sigma$}         &  
 \colhead{Fe EW}             &\colhead{$A_{\rm ga}$$^b$}   & 
 \colhead{$A_{\rm scat}$$^c$}     &\colhead{$\chi^2_{\rm \nu}$(dof)} \\ 
                             & \colhead{($10^{22}$ cm$^{-2}$)} &
                             &                               &
 \colhead{(keV)}             & \colhead{(eV)}                & 
  \colhead{(eV)}             &                               &
                             &
}               
\startdata
 Mrk 348        & 19.07$^{+0.56}_{-0.52}$ & 1.677$^{+0.021}_{-0.018}$ & 11.89$^{+0.20}_{-0.21}$ & 6.448$^{+0.026}_{-0.032}$  &  5($<$69)          &52$^{+18}_{-12}$    & 2.60$^{+0.86}_{-0.68}$  & 3.41$^{+0.46}_{-0.31}$ & 1.10(137) \\  
 3C 33          & 77$^{+22}_{-16}$        & 2.03$^{+0.23}_{-0.25}$   & 6.3$^{+0.9}_{-2.7}$     & 6.41$^{+0.10}_{-0.07}$     & 55($<$149)          &180$\pm120$        & 2.8$\pm1.8$           & 1.22$^{+0.56}_{-0.62}$    & 1.13(40)  \\ 
 NGC 1142       & 56.5$^{+3.2}_{-3.6}$    &1.815$^{+0.047}_{-0.073}$  & 5.19$\pm0.26$         &6.414$\pm0.028$          & 20($<$69)            &220$\pm60$         & 3.8$\pm1.2$           & 2.84$\pm0.31$        & 1.16(90)   \\  
 3C 98          & 12.3$^{+1.3}_{-0.8}$    & 1.64$^{+0.07}_{-0.10}$   & 1.11$^{+0.24}_{-0.27}$  &6.4(f)                   & 0(f)              &65$\pm$50          & 0.36$\pm0.27$          & 1.20$^{+0.39}_{-0.28}$ & 1.01(142)   \\ 
 IC 2461        & 6.84$^{+0.20}_{-0.18}$  & 1.691$^{+0.048}_{-0.051}$  & 1.305$^{+0.028}_{-0.025}$&6.409$\pm0.034$          & 105$^{+47}_{-36}$    &211$^{+47}_{-49}$    & 1.20$^{+0.27}_{-0.28}$   & 0.18$^{+0.10}_{-0.11}$ & 0.88(116) \\
 NGC 4138       & 7.60$^{+0.40}_{-0.38}$  & 1.492$^{+0.055}_{-0.057}$  & 1.460$^{+0.041}_{-0.045}$&6.380$^{+0.068}_{-0.041}$ &  10(f)              &86$^{+44}_{-47}$     & 0.79$^{+0.44}_{-0.43}$   & 1.17$^{+0.23}_{-0.24}$ & 1.03(178) \\ 
 NGC 4388       & 29.1$^{+1.2}_{-1.1}$   & 1.479$^{+0.036}_{-0.035}$  & 8.59$^{+0.16}_{-0.17}$  & 6.431$\pm0.016$        & 42($<$68)           &220$\pm30$          & 12.0$^{+2.0}_{-1.8}$    & 4.67$^{+0.71}_{-0.65}$ & 1.10(157)  \\  
 NGC 4507       & 42.84$^{+0.81}_{-0.78}$  & 1.500$\pm0.012$        & 7.050$^{+0.079}_{-0.088}$  & 6.419$\pm0.013$      & 5($<$34)             &120$\pm16$         &5.80$^{+0.77}_{-0.74}$    & 9.81$\pm0.32$ & 1.17(199) \\
 ESO 506$-$G027 & 117$^{+14}_{-11}$      & 1.557$^{+0.072}_{-0.093}$  & 6.8$^{+3.2}_{-0.3}$     & 6.408$^{+0.024}_{-0.022}$ & 51$^{+30}_{-40}$      & 520$^{+130}_{-170}$&20.5$^{+5.1}_{-6.6}$      & 1.25$\pm0.19$        & 0.93(79) \\
 NGC 4939       & 18.6$^{+2.9}_{-2.4}$   & 1.9(f)                   & 3.03$^{+0.51}_{-0.45}$     & 6.4(f)                 & 10(f)               &87($<$220)       & 0.84($<$2.02)          & 2.85$^{+0.69}_{-0.77}$    & 1.08(27) \\ 
 ESO 383$-$G18  & 18.75$^{+0.98}_{-0.96}$ & 1.440$^{+0.063}_{-0.064}$  & 1.88$^{+0.12}_{-0.09}$  &6.4(f)                  & 30($<$130)           &100$^{+50}_{-40}$   & 1.34$^{+0.62}_{-0.55}$   & 1.47$^{+0.24}_{-0.27}$  & 1.11(133)\\ 
 ESO 103$-$G035 & 20.60$^{+0.57}_{-0.56}$ & 1.922$^{+0.039}_{-0.041}$  & 19.14$^{+0.28}_{-0.29}$ &6.475$\pm0.026$         & 3($<$43)             & 66$^{+22}_{-19}$   & 3.6$\pm1.1$           & 1.30$^{+0.35}_{-0.33}$ & 1.02(133)   \\
 IC 4995        & 45$^{+21}_{-9}$       & 1.9(f)                   & 0.45$^{+0.10}_{-0.16}$  & 6.390$^{+0.057}_{-0.036}$ & 0($<$66)             &890$^{+420}_{-440}$ & 1.20$^{+0.56}_{-0.59}$  & 0.90$^{+0.35}_{-0.32}$   &   1.14(15)\\ 
 NGC 7172       & 7.872$\pm0.071$       & 1.661$\pm0.014$          & 15.78$\pm0.09$        & 6.380$\pm0.020$        & 72$^{+28}_{-24}$      & 69$\pm11$        & 5.10$\pm0.74$           & 2.81$^{+0.24}_{-0.33}$ & 1.18(168)    \\  
 NGC 7319       & 75.5$^{+6.2}_{-5.7}$   & 1.90$^{+0.12}_{-0.26}$    & 2.70$^{+0.11}_{-0.87}$   & 6.392$\pm0.023$         & 54($<$97)           &370$\pm80$        &3.00$^{+0.65}_{-0.62}$     & 1.25$\pm0.20$        & 1.13(128) 
\enddata
\tablecomments{Photon index of the power law with only Galactic absorption was assumed to be the same value as the hard power law absorbed by cold matter at the redshift of the source. (f) indicates fixed parameter.}
\tablenotetext{a}{Normalization of the absorbed power law in units of $10^{-3}$photons cm$^{-2}$ s$^{-1}$ at 1 keV.}
\tablenotetext{b}{Normalization of the Gaussian line in units of $10^{-5}$ photons cm$^{-2}$ s$^{-1}$ in the line.}
\tablenotetext{c}{Normalization of the less absorbed power law in units of $10^{-5}$ photons cm$^{-2}$ s$^{-1}$ at 1 keV.}
\label{table:complex}
\end{deluxetable*}
}

{\tabletypesize{\scriptsize}
\begin{deluxetable*}{lccccccc}
%\tablewidth{\columnwidth}
%\tablecolumns{8}
\tablecaption{Spectral Parameters for MEKAL and the Third Power Law in the Complex Models}
\tablewidth{0pt}
\tablehead{
\colhead{}\\    
               &\multicolumn{2}{c}{MEKAL$^a$} &&\multicolumn{2}{c}{MEKAL$^a$} &&\multicolumn{1}{c}{Power law}\\ \cline{2-3}\cline{5-6}\cline{8-8}\\
\colhead{Name} &\colhead{\it kT}    & \colhead{$A_{\rm m}$$^b$}    &&\colhead{\it kT} & \colhead{$A_{\rm m}$$^b$} &\colhead{$N_{\rm H}$$^c$}   & \colhead{$A^d$}       \\
               &\colhead{(keV)} &                       &&\colhead{(keV)} &                     &\colhead{($10^{22}$ cm$^{-2}$)} &                      \\
}               
\startdata
 Mrk 348       & 0.68$^{+0.13}_{-0.09}$& 2.04$\pm0.38$       && 0.169$^{+0.015}_{-0.013}$ & 6.0$\pm1.0$ & 7.48$\pm0.20$ & 2.78$\pm0.14$ \\  
 3C 33         & 0.29$^{+0.10}_{-0.07}$   & 2.07$^{+0.91}_{-0.87}$&&\nodata&\nodata&\nodata&\nodata\\ 
 NGC 1142      & 0.295$^{+0.033}_{-0.043}$& 3.70$^{+0.72}_{-0.79}$&&\nodata&\nodata&\nodata&\nodata\\  
 3C 98         & 0.65(f)                & 1.69$^{+0.54}_{-0.43}$&&\nodata&\nodata&\nodata&\nodata\\ 
 IC 2461       & 0.211$^{+0.054}_{-0.049}$& 0.39$^{+0.17}_{-0.15}$&&\nodata&\nodata&\nodata&\nodata\\
 NGC 4138      & 0.265$^{+0.062}_{-0.048}$& 1.44$\pm$0.42       &&\nodata&\nodata&\nodata&\nodata\\  
 NGC 4388      & 0.78$^{+0.11}_{-0.10}$   & 7.60$^{+0.89}_{-0.93}$&& 0.203$^{+0.056}_{-0.014}$ & 10.9$^{+0.7}_{-1.6}$ & 2.5$^{+1.4}_{-1.0}$  &\nodata   \\  
 NGC 4507      & \nodata                & \nodata             &&\nodata&\nodata&9.1$^{+3.0}_{-2.5}$ &\nodata\\ 
 ESO 506$-$G027& \nodata                & \nodata             &&\nodata&\nodata& 12.5$^{+3.0}_{-2.5}$ & \nodata\\
 NGC 4939      & 0.65(f)                & 1.9$^{+1.0}_{-1.2}$  &&\nodata&\nodata&\nodata&\nodata\\ 
 ESO 383$-$G18 & 0.232$^{+0.030}_{-0.036}$& 2.77$^{+0.63}_{-0.57}$&&\nodata&\nodata&\nodata&\nodata\\  
 ESO 103$-$G035& 0.81$^{+0.16}_{-0.11}$   & 2.27$^{+0.50}_{-0.48}$&&\nodata&\nodata&2.42$^{+0.73}_{-0.49}$& \nodata\\  
 IC 4995       & 0.52$^{+0.12}_{-0.14}$   & 2.40$^{+0.57}_{-0.55}$&&0.081($<$0.085)& 113$^{+21}_{-24}$&\nodata&\nodata\\
 NGC 7172      & 0.302$^{+0.073}_{-0.061}$& 1.30$\pm0.37$       &&\nodata&\nodata&\nodata&\nodata\\  
 NGC 7319      & 0.621$^{+0.034}_{-0.037}$& 3.51$\pm0.36$       &&\nodata&\nodata& 7.7$^{+2.8}_{-2.2}$ & \nodata
\enddata
\tablecomments{(f) indicates the fixed parameter.}
\tablenotetext{a}{Metal abundances were fixed at 0.5 solar.}
\tablenotetext{b}{Normalization of {\tt mekal} in units of 10$^{-19}$/(4$\pi$({\it D}$_{\rm A}\times$(1+{\it z}))$^2$) $\int$ {\it n}$_{\rm e}$ {\it n}$_{\rm H}$ {\it dV}, where {\it D}$_{\rm A}$ is the angular size distance to the source (cm), {\it n}$_{\rm e}$ is the electron density (cm$^{-3}$), and {\it n}$_{\rm H}$ is the hydrogen density (cm$^{-3}$).}
\tablenotetext{c}{Absorption of pexrav or the third power law by cold matter at the redshift of the source.}
\tablenotetext{d}{Normalization of the third power law in units of $10^{-3}$photons cm$^{-2}$ s$^{-1}$ at 1 keV .}
\end{deluxetable*}}

\newcounter{ichi}
\setcounter{ichi}{1}
\newcounter{nana}
\setcounter{nana}{7}
\newcounter{hati}
\setcounter{hati}{8}
\newcounter{juunana}
\setcounter{juunana}{17}
\newcounter{kyuu}
\setcounter{kyuu}{9}
\newcounter{juu}
\setcounter{juu}{10}
\newcounter{juuichi}
\setcounter{juuichi}{11}
\newcounter{juusan}
\setcounter{juusan}{13}
\newcounter{juuroku}
\setcounter{juuroku}{16}
\newcounter{nijuusan}
\setcounter{nijuusan}{23}
\newcounter{nijuugo}
\setcounter{nijuugo}{25}
\newcounter{nijuuroku}
\setcounter{nijuuroku}{26}

{\tabletypesize{\scriptsize}
\begin{deluxetable}{lcccc}
\tablewidth{\columnwidth}
%\tablecolumns{5}
\tablecaption{Spectral Parameters for Gaussians in the Complex Models}
%\setlength{\tabcolsep}{-3pt}
%\tablewidth{0pt}
\tablehead{
\colhead{}\\    
 \colhead{Name}       &\colhead{$E_{\rm line}$} &\colhead{$\sigma$}    &\colhead{$A_{\rm ga}$$^a$} &\colhead{Identification} \\  
                      &\colhead{(keV)}         & (eV)                &                           &                             \\
}               
\startdata
Mrk 348.............  &6.670$\pm0.045$         & 60($<$111)           &$-$2.52$^{+0.72}_{-0.60}$   & Fe\ \Roman{nijuugo}  \\
                      &7.000$^{+0.035}_{-0.032}$ & 0($<$67)             &$-$1.74$^{+0.59}_{-0.54}$& Fe\ \Roman{nijuuroku}\\
&&&&\\
NGC 7172..........    &1.734$^{+0.046}_{-0.041}$ & 77$^{+51}_{-36}$      & 0.57$\pm0.20$        & Si\ K$\alpha$ \\
&&&&\\
NGC 4507..........    &0.475$\pm0.007$         & 10(f)                & 6.50$^{+0.78}_{-0.75}$ & O\ \Roman{nana}\ K$\alpha$ \\
                      &0.560$^{+0.003}_{-0.004}$ & 10(f)                & 11.00$^{+0.79}_{-0.76}$ & O\ \Roman{nana}\ K$\alpha$ \\
                      &0.648$\pm0.007$         & 10(f)                & 3.15$^{+0.42}_{-0.40}$ & O\  \Roman{hati}\ K$\alpha$\\
                      &0.758$\pm0.006$         & 46$^{+7}_{-6}$        & 5.04$\pm0.37$       & O\  \Roman{nana}\ RRC\\ 
                      &0.904$\pm0.005$         & 10(f)                & 3.31$^{+0.26}_{-0.24}$ & Ne\ \Roman{kyuu}\ K$\alpha$\\
                      &1.013$^{+0.014}_{-0.013}$ & 10(f)                & 0.93$\pm0.17$        & Ne\ \Roman{juu}\ K$\alpha$\\
                      &1.169$^{+0.016}_{-0.018}$ & 10(f)                & 0.52$\pm0.12$ & Fe\ \Roman{nijuusan}\ L\\
                      &1.330$^{+0.012}_{-0.013}$ & 10(f)                & 0.53$^{+0.09}_{-0.13}$ & Mg\ \Roman{juuichi}\ K$\alpha$\\
                      &1.788$^{+0.033}_{-0.026}$ & 10(f)                & 0.204$\pm0.085$& Si\ \Roman{juusan}\ K$\alpha$\\
                      &6.320(f)                & 40(f)                 & 3.05 $^{+0.79}_{-0.77}$ & Fe\ \Roman{ichi}$-$\Roman{juuroku}\ CS$^{*b}$   
\enddata
\tablecomments{(f) indicates fixed the parameter.}
\tablenotetext{a}{Normalization of the Gaussian line in units of $10^{-5}$ photons cm$^{-2}$ s$^{-1}$.}
\tablenotetext{b}{Compton shoulder.}
\end{deluxetable}
}

{\tabletypesize{\scriptsize}
\begin{deluxetable}{ll}
\tablewidth{\columnwidth}
%\tablecolumns{2}
\tablecaption{Best-Fit Models for our Sample}
%\setlength{\tabcolsep}{-3pt}
%\tablewidth{0pt}
\tablehead{\colhead{Name}& \colhead{Model$^a$}
}
\startdata
  Mrk 348                   & BM + two MEKAL + two lines + abs-PL  \\
  3C 33                     & BM + MEKAL + Ref\\
  2MASX J02281350$-$0315023 & BM   \\
  NGC 1142                  & BM + MEKAL   \\
  3C 98                     & BM + MEKAL  \\
  B2 0857+39                & BM   \\
  IC 2461                   & BM + MEKAL  \\
  2MASX J10335255+0044033   & BM   \\
  MCG +08-21-065            & BM   \\
  NGC 4074                  & BM   \\ 
  NGC 4138                  & BM + MEKAL  \\
  NGC 4388                  & BM + two MEKAL + abs-Ref\\
  NGC 4507                  & BM + ten lines + abs-Ref \\
  ESO 506-G027              & BM + abs-Ref  \\
  2MASX J12544196$-$3019224 & BM   \\
  NGC 4939                  & BM + MEKAL  \\
  ESO 383-G18               & BM + MEKAL + Ref \\
  ESO 103-G035              & BM + MEKAL + abs-Ref  \\
  IC 4995                   & BM + two MEKAL  \\
  NGC 7070A                 & BM   \\
  NGC 7172                  & BM + MEKAL + line \\
  NGC 7319                  & BM + MEKAL + abs-Ref  
\enddata
\tablenotetext{a}{BM: Baseline model, PL: power law, MEKAL: thin thermal plasma model ({\tt mekal}), Ref: cold refection model ({\tt pexrav}), abs-PL(or Ref): absorbed PL(or Ref). All components are absorbed by the Galactic absorption.}
\label{table:model}
\end{deluxetable}}

{\tabletypesize{\scriptsize}
\begin{deluxetable*}{lccccccccccccc}
\addtolength{\tabcolsep}{-3pt}
\tablecolumns{14}
\tablecaption{Flux for the Best-Fit Model}
\tablewidth{0pt}
\tablehead{
\colhead{}\\    
  &\multicolumn{2}{c}{Total}               &&  \multicolumn{4}{c}{Power Law$_{\rm total}$}  &&
   \multicolumn{2}{c}{Power Law$_{\rm scat}$}&&  \multicolumn{2}{c}{MEKAL}\\ \cline{2-3}\cline{5-8}\cline{10-11}\cline{13-14}\\
   \colhead{Name}        &\colhead{(Observed)} &\colhead{(Observed)}&&  
   \colhead{(Observed)}  &\colhead{(Observed)} &\colhead{(Intrinsic)}&
   \colhead{(Intrinsic)}&&\colhead{(Observed)} &\colhead{(Intrinsic)}&&
   \colhead{(Observed)}  &\colhead{(Intrinsic)}\\
  &\colhead{0.5$-$2 keV}    &\colhead{2$-$10 keV}    &&\colhead{0.5$-$2 keV} &
   \colhead{2$-$10 keV}     &\colhead{0.5$-$2 keV}   &\colhead{2$-$10 keV}   &&
   \colhead{0.5$-$2 keV}    &\colhead{0.5$-$2 keV}   &&\colhead{0.5$-$2 keV} & 
   \colhead{0.5$-$2 keV}\\
   \colhead{(1)}&\colhead{(2)}&\colhead{(3)}&&
   \colhead{(4)}&\colhead{(5)}&\colhead{(6)}&
   \colhead{(7)}&&\colhead{(8)}&\colhead{(9)}&&
   \colhead{(10)}&\colhead{(11)}\\
}
\startdata                       
  Mrk 348                   &12.3   &27.9    &&8.36    &28.1    &32.2   &60.9  &&6.33   &7.49   &&3.97   &5.50   \\  %5.31     23.0
  3C 33                     &6.06   &2.51    &&2.16    &1.56    &12.4   &13.8  &&2.16   &2.41   &&1.41   &1.69   \\  %1.81
  2MASX J02281350$-$0315023 &0.637  &0.817   &&0.637   &0.792   &1.44   &1.94  &&0.634  &0.681  &&\nodata&\nodata\\         
  NGC 1142                  &7.49   &3.11    &&5.04    &2.97    &11.1   &17.0  &&5.03   &6.01   &&2.45   &3.35   \\ %2.20
  3C 98                     &3.49   &2.62    &&2.02    &2.59    &2.40   &4.82  &&1.94   &2.57   &&1.47   &2.14   \\ %2.56
  B2 0857+39                &0.376  &0.420   &&0.376   &0.417   &0.624  &0.841 &&0.353  &0.377  &&\nodata&\nodata\\
  IC 2461                   &1.92   &3.55    &&1.67    &3.44    &2.89   &5.35  &&0.383  &0.396  &&0.254  &0.273  \\  %0.280
  2MASX J10335255+0044033   &1.10   &0.557   &&1.10    &0.557   &0.918  &1.24  &&1.10   &1.28   &&\nodata&\nodata\\
  MCG +08-21-065            &0.415  &0.843   &&0.415   &0.814   &1.71   &2.21  &&0.414  &0.439  &&\nodata&\nodata\\ 
  NGC 4074                  &6.17   &2.75    &&6.17    &2.66    &4.90   &6.61  &&6.16   &6.64   &&\nodata&\nodata\\ 
  NGC 4138                  &4.77   &5.44    &&3.58    &5.37    &3.32   &8.30  &&2.56   &2.65   &&1.19   &1.28   \\  %1.26
  NGC 4388                  &25.2   &20.7    &&9.73    &17.1    &19.4   &49.4  &&9.73   &10.5   &&14.9   &16.9   \\ %17.3
  NGC 4507                  &36.6   &12.5    &&18.1    &10.2    &15.9   &39.1  &&18.1   &21.9   &&\nodata&\nodata\\  %17.6
  ESO 506-G027              &2.30   &4.06    &&2.30    &2.22    &15.1   &34.2  &&2.30   &2.68   &&\nodata&\nodata\\
  2MASX J12544196$-$3019224 &1.05   &0.527   &&1.05    &0.491   &0.906  &1.22  &&1.05   &1.27   &&\nodata&\nodata\\
  NGC 4939                  &7.79   &3.65    &&5.60    &3.59    &6.66   &8.98  &&5.60   &6.22   &&2.19   &2.47   \\
  ESO 383-G18               &5.42   &6.09    &&2.94    &5.25    &4.24   &11.5  &&2.94   &3.28   &&1.66   &2.14   \\ %1.79
  ESO 103-G035              &5.72   &23.2    &&2.37    &20.1    &41.5   &54.3  &&2.35   &2.82   &&2.40   &2.90   \\  %2.43
  IC 4995                   &6.29   &0.348   &&1.71    &0.295   &0.992  &1.34  &&1.71   &1.94   &&4.59   &5.92   \\  %6.42
  NGC 7070A                 &1.43   &2.17    &&1.43    &2.16    &1.63   &3.99  &&1.39   &1.50   &&\nodata&\nodata\\
  NGC 7172                  &17.5   &42.3    &&14.8    &41.8    &34.9   &67.6  &&5.85   &6.21   &&1.13   &1.25   \\ %2.58
  NGC 7319                  &5.80   &1.34    &&2.16    &0.905   &5.78   &7.82  &&2.16   &2.61   &&3.62   &4.56    %3.50
\enddata
\tablecomments{The units of Columns 2, 4, 8, 9, 10, and 11 are $10^{-14}$ erg cm$^{-2}$ s$^{-1}$. The units of Columns 3, 5, 6, and 7 are $10^{-12}$ erg cm$^{-2}$ s$^{-1}$.}
\label{table:flux}
\end{deluxetable*}
}

{\tabletypesize{\scriptsize}
\begin{deluxetable}{lccccc}
\tablecaption{Scattering Fraction}
%\tablecolumns{6}
\tablewidth{\columnwidth}
\tablecaption{Hard X-ray, Soft X-ray, Far Infrared, and [\ion{O}{3}] $\lambda$5007 Luminosities}
\tablehead{
\colhead{Name}  &
\colhead{Hard}  & \colhead{Soft}     & 
\colhead{FIR}   & \colhead{[\ion{O}{3}]}  & \colhead{Reference} \\
\\
\colhead{(1)}&\colhead{(2)}&\colhead{(3)}&\colhead{(4)}&\colhead{(5)}&\colhead{(6)}\\
}
\startdata
  Mrk 348                   & 43.49 & 40.82 &  43.49    &  41.95      & 1\\
  3C 33                     & 44.07 & 41.55 &  \nodata  &  42.52      & 2\\
  2MASX J02281350$-$0315023 & 43.46 & 41.01 &  \nodata  &  \nodata    & \nodata\\
  NGC 1142                  & 43.51 & 41.25 &  44.77    &  41.87      & 3\\
  3C 98                     & 43.01 & 41.00 &  \nodata  &  41.91      & 4\\
  B2 0857+39                & 44.12 & 41.77 &  \nodata  &  \nodata    & \nodata\\
  IC 2461                   & 41.83 & 38.93 &  43.00    &  \nodata    & \nodata\\
  2MASX J10335255+0044033   & 43.75 & 41.77 &  \nodata  &  43.62      & 5\\
  MCG +08-21-065            & 42.65 & 39.95 &  43.51    &  39.78      & 6\\
  NGC 4074                  & 42.88 & 40.88 &  \nodata  &  42.05      & 7\\ 
  NGC 4138                  & 41.21 & 38.88 &  \nodata  &  38.75      & 8\\
  NGC 4388                  & 42.89 & 40.63 &  43.94    &  41.77      & 1\\
  NGC 4507                  & 43.09 & 41.19 &  43.81    &  41.69      & 1\\
  ESO 506-G027              & 43.69 & 40.58 &  43.58    &  42.96      & 9\\
  2MASX J12544196$-$3019224 & 42.96 & 40.98 &  \nodata  &  \nodata    & \nodata\\
  NGC 4939                  & 42.33 & 40.31 &  43.59    &  41.43      & 1\\
  ESO 383-G18               & 42.60 & 40.27 &  \nodata  &  \nodata    & \nodata\\
  ESO 103-G035              & 43.33 & 40.35 &  43.54    &  41.65      & 1\\
  IC 4995                   & 41.88 & 40.66 &  43.40    &  41.96      & 7\\
  NGC 7070A                 & 41.75 & 39.33 &  42.40    &  \nodata    & \nodata\\
  NGC 7172                  & 43.06 & 40.18 &  43.77    &  39.83      & 1\\
  NGC 7319                  & 42.95 & 40.92 &  \nodata  &  41.44      & 1
\enddata
\label{table:flux_OIII_FIR}
\tablecomments{Column 1: Galaxy name. Column 2: Logarithm of intrinsic 2$-$10 keV luminosity. Column 3: Logarithm of intrinsic 0.5$-$2 keV luminosity of soft X-ray components. Column 4: Logarithm of far-infrared luminosity. Column 5: Logarithm of reddening corrected [\ion{O}{3}] luminosity. Column 6: References for [\ion{O}{3}] luminosity.}
\tablerefs{(1) Bassani et al. 1999; (2) Yee \& Oke 1978; (3) Shu et al. 2007; (4) Costero \& Osterbrock 1977; (5) Dong et al. 2005; (6) Line flux measurement based on the Sloan Digital Sky Survey data at MPA/JHU (http://www.mpa-garching.mpg.de/SDSS/); Kauffman et al. 2003; (7) Polletta et al. 1996; (8) Ho et al. 1997; (9) Landi et al. 2007.}
\end{deluxetable}
}

\section{Discussion}
\subsection{The Origin of the Soft X-ray Emission}
\label{FIR}

Since various spectral components presumably contribute to the soft
X-ray emission in obscured AGNs, understanding the origin of the soft
emission is of importance to derive true scattering fractions. One of
the possible origins is the circumnuclear gas photoionized and
photoexcited by AGN emission. The soft X-ray emission is also produced
through Thomson scattering of the primary radiation by free electrons
in the ionized gas. High-resolution spectra of some Seyfert 2 galaxies
obtained with {\it Chandra} or {\it XMM-Newton} have shown that
emission from a photoionized or photoexcited plasma is dominant in soft
X-rays (e.g., Sako et al. 2000; Kinkhabwala et al. 2002). Another
possibility is the contribution of thermal emission from a collisionally
ionized plasma, which is heated by shocks induced by AGN outflows
(King 2005) or intense star formation.  For example, a high-resolution
image of NGC 4945 with {\it Chandra} suggested that the soft X-rays
are mostly dominated by thermal radiation from a starburst region
(Schurch et al. 2002).

\begin{figure}[!htb]
\centering
\includegraphics[angle=270,scale=.40]{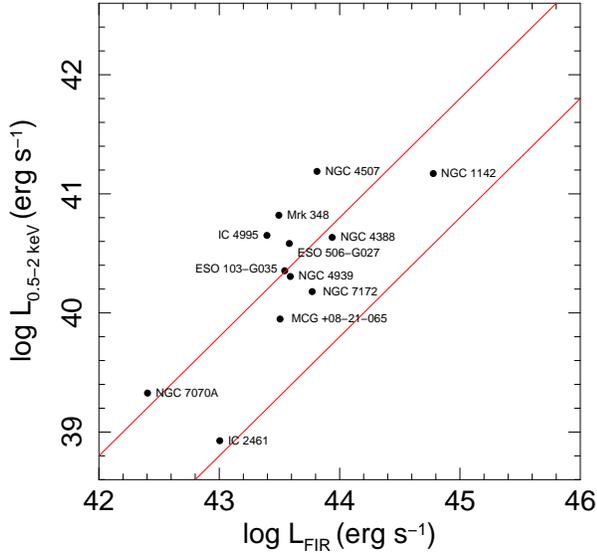}
\caption{Comparison between intrinsic luminosity of the soft component in
  the 0.5$-$2 keV band and FIR luminosity. The area surrounded by two solid lines is a
  typical region for starburst galaxies (Ranalli et al. 2003).}
\label{figure:FIR_X}
\end{figure}

Far-infrared (FIR) luminosities are often used to estimate the
contribution of starburst since star-forming activity is much more
effective than the AGN in powering the FIR emission. A tight linear
relation between the soft X-ray and the FIR luminosity is known for
starburst and normal galaxies (e.g., David et al. 1992; Ranalli et
al. 2003). Thus, the FIR luminosity can be used to determine the
contribution of starburst to the soft X-ray. The FIR luminosities of
our sample were calculated using the formula defined in Helou et
al. (1985), based on flux densities at 60 $\mu$m and 100
$\mu$m. Infrared fluxes (60 $\mu$m and 100 $\mu$m) measured with {\it
  Infrared Astronomical Satellite} ({\it IRAS}) for 12 objects were
taken from the NED. {\it IRAS} fluxes for the rest of the 10 objects
are not available. The soft X-ray luminosities were calculated from
fluxes in 0.5$-$2 keV corrected for absorption given in Table 7,
Columns 9 and 11, except for NGC 4507 and NGC 7172. For
calculations of the soft X-ray luminosities of NGC 4507 and NGC 7172,
absorption corrected fluxes of the Gaussians in 0.5$-$2 keV (NGC 4507:
27.6$\times$10$^{-14}$ erg cm$^{-2}$ s$^{-1}$, NGC 7172:
1.53$\times$10$^{-14}$ erg cm$^{-2}$ s$^{-1}$), were added to the
fluxes given in Table 7, Columns 9 and 11. The calculated FIR and
soft X-ray luminosities are shown in Table \ref{table:flux_OIII_FIR}
and the relation between the soft X-ray and the FIR luminosity for our
sample is shown in Figure \ref{figure:FIR_X}. The area surrounded by
two solid lines expresses the region for starburst galaxies in Ranalli
et al. (2003). About a half of our sample is in the starburst
region. This means that the contribution from starburst is likely to
be large in the soft X-ray emission of these sources. Therefore, it
should be noted that starburst contribution may not be negligible in
the soft X-ray emission in calculating scattering fractions in section
4.2 for about a half of our sample. If starburst significantly
contributes to the soft X-ray emission, the level of scattered
emission would be much lower than those derived in Section 4.2.

\subsection{Scattering Fraction}
\label{scattering fraction}
Scattering fractions are often calculated with the following equation;
\begin{displaymath}
{\it f}_{\rm scat}=\frac{A_{\rm scat}}{A_{\rm int}}, 
\end{displaymath}
where $A_{\rm int}$ and $A_{\rm scat}$ are the
normalization for the power law with large and only Galactic absorption,
respectively. The values calculated for our sample are shown in Table
\ref{table:Fs}. Since the origin of soft X-ray emission is not clear
as discussed in Section \ref{FIR}, we calculated scattering fractions with
the equation as follows;
\begin{displaymath}
 {\it f}_{\rm scat}^{\rm obs}=\frac{F_{0.5-2}^{\rm soft}}{F_{0.5-2}^{\rm int}} , 
\end{displaymath}
where $F_{0.5-2}^{\rm int}$ and $F_{\rm 0.5-2}^{\rm soft}$ are
absorption corrected fluxes in the 0.5$-$2 keV band for the power law
corresponding to the direct emission and all the components except for
the direct power law, respectively. This value is regarded as an upper
limit on the scattering fraction. The obtained values (Table
\ref{table:Fs}) except for NGC 4507 and IC 4995 are smaller than the
value typical for other Seyfert 2s observed so far, which is about 3\%
(e.g., Turner et al. 1997; Bianchi and Guainazzi 2007). In particular,
those of eight sources are less than 0.5\%. These are in the range
recently found by hard X-ray surveys (Ueda et al. 2007; Winter et
al. 2008).

The scattering fraction can be related to the geometry of the
scatterer;
\begin{displaymath}
 {\it f}_{\rm scat}\sim\tau\frac{\Delta\Omega}{4\pi},
\label{scat_f}
\end{displaymath}
where $\Delta\Omega$ and $\tau$ are the solid angle subtended by the
scattering electrons and a scattering optical depth,
respectively. Thus, the small scattering fraction indicates that
$\Delta\Omega$ and/or $\tau$ are small. If $\tau$ does not differ much
from object to object, our sample with a small scattering fraction is
strong candidates for AGNs buried in a very geometrically thick torus
with a small opening angle.

{\tabletypesize{\scriptsize}
\begin{deluxetable}{lcc}
%\tablecolumns{3}
\tablecaption{Scattering Fraction}
\tablewidth{\columnwidth}
\tablehead{\colhead{Name}& \colhead{$f_{\rm scat}$$^a$} & \colhead{$f_{\rm scat}^{\rm obs}$$^b$}\\
\colhead{}& \colhead{(\%)} & \colhead{(\%)}\\
}
\startdata
  Mrk 348                   & 0.29 &  0.40   \\
  3C 33                     & 0.19 &  0.33   \\
  2MASX J02281350$-$0315023 & 0.47 &  0.47   \\
  NGC 1142                  & 0.55 &  0.84   \\
  3C 98                     & 1.1  &  2.0    \\
  B2 0857+39                & 0.61 &  0.61   \\
  IC 2461                   & 0.14 &  0.23   \\
  2MASX J10335255+0044033   & 1.4  &  1.4    \\
  MCG +08-21-065            & 0.26 &  0.26   \\
  NGC 4074                  & 1.4  &  1.4    \\ 
  NGC 4138                  & 0.80 &  1.2    \\
  NGC 4388                  & 0.54 &  1.4    \\
  NGC 4507                  & 1.4  &  3.1    \\
  ESO 506-G027              & 0.18 &  0.18   \\
  2MASX J12544196$-$3019224 & 1.4  &  1.4    \\
  NGC 4939                  & 0.94 &  1.3    \\
  ESO 383-G18               & 0.78 &  1.3    \\
  ESO 103-G035              & 0.068&  0.14   \\
  IC 4995                   & 2.0  &  8.1    \\
  NGC 7070A                 & 0.93 &  0.93   \\
  NGC 7172                  & 0.18 &  0.26   \\
  NGC 7319                  & 0.46 &  1.3    
\enddata
\tablenotetext{a}{Scattering fraction calculated by {\it f}$_{\rm scat}$=$\frac{A_{\rm scat}}{A_{\rm int}}$, where $A_{\rm int}$ and $A_{\rm scat}$ are the normalization for the power law with large and only Galactic absorption, respectively.}
\tablenotetext{b}{Scattering fraction calculated by {\it f}$_{\rm scat}^{\rm obs}$=$\frac{F_{0.5-2}^{\rm soft}}{F_{0.5-2}^{\rm int}}$, where $F_{0.5-2}^{\rm int}$ and $F_{\rm 0.5-2}^{\rm soft}$ are absorption corrected fluxes in the 0.5$-$2 keV band for the power law corresponding to the direct emission and all the components except for the direct power law, respectively.}
\label{table:Fs}
\end{deluxetable}
}

\subsection{Comparison with [\ion{O}{3}]$\lambda$5007 luminosity}
[\ion{O}{3}] $\lambda$5007 emission is produced in the narrow line
region (NLR), which is considered to exist in the opening direction of
the torus. Since the X-ray scattering region is also spatially
extended along the NLR (Sako et al. 2000; Young et al. 2001; Bianchi
et al. 2006), we expect that an AGN buried in a geometrically thick
torus with a small opening part should have a fainter [\ion{O}{3}]
emission luminosity relative to a hard X-ray luminosity compared with
classical Seyfert 2 galaxies with a large opening part.

We collected [\ion{O}{3}] luminosities for 16 objects in our sample
from the literature as shown in Table \ref{table:flux_OIII_FIR}. The
[\ion{O}{3}] fluxes were corrected for the extinction by using the
relation
\begin{displaymath}
 {\it L}_{\rm [O\ III]}^{\rm int}={\it L}_{\rm [O\ III]}^{\rm obs}\biggl[\frac{{\rm
       H}\alpha/{\rm H}\beta}{({\rm H}\alpha/{\rm
       H}\beta)_0}\biggr]^{2.94},
\end{displaymath}
assuming an intrinsic Balmer decrement (H$\alpha$/H$\beta$)$_0$ = 3.0,
where $L_{[\rm O\ III]}^{\rm obs}$ and {\rm H}$\alpha/${\rm H}$\beta$
are an observed [\ion{O}{3}] luminosity and a ratio between observed
{\rm H}$\alpha$ and {\rm H}$\beta$ line fluxes, respectively (Bassani
et al. 1999).

\begin{figure}
\centering
\includegraphics[angle=270,scale=.40]{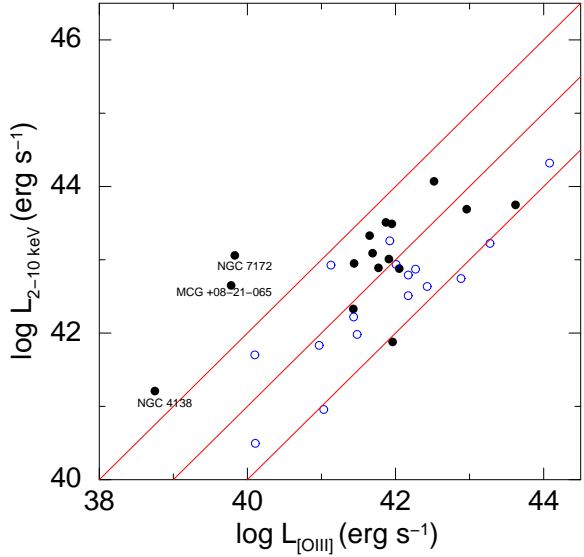}
\caption{Comparison between intrinsic luminosity in the 2$-$10 keV
  band and reddening corrected [\ion{O}{3}] line luminosity. Our
  sample is shown as filled circles. Other Seyfert 2s with {\it
    N}$_{\rm H}$ in the range of (0.6$-$20)$\times$10$^{23}$ cm$^{\rm
    -2}$ taken from Bassani et al. (1999) are shown as open
  circles. Solid lines correspond to {\it L}$_{\rm 2-10 keV}$/{\it L}$_{[{\rm
        O\ III}]}$ = 1, 10, and 100 from bottom to top.}
\label{figure:OIII}
\end{figure}

\begin{figure}
\centering
\includegraphics[angle=270,scale=.40]{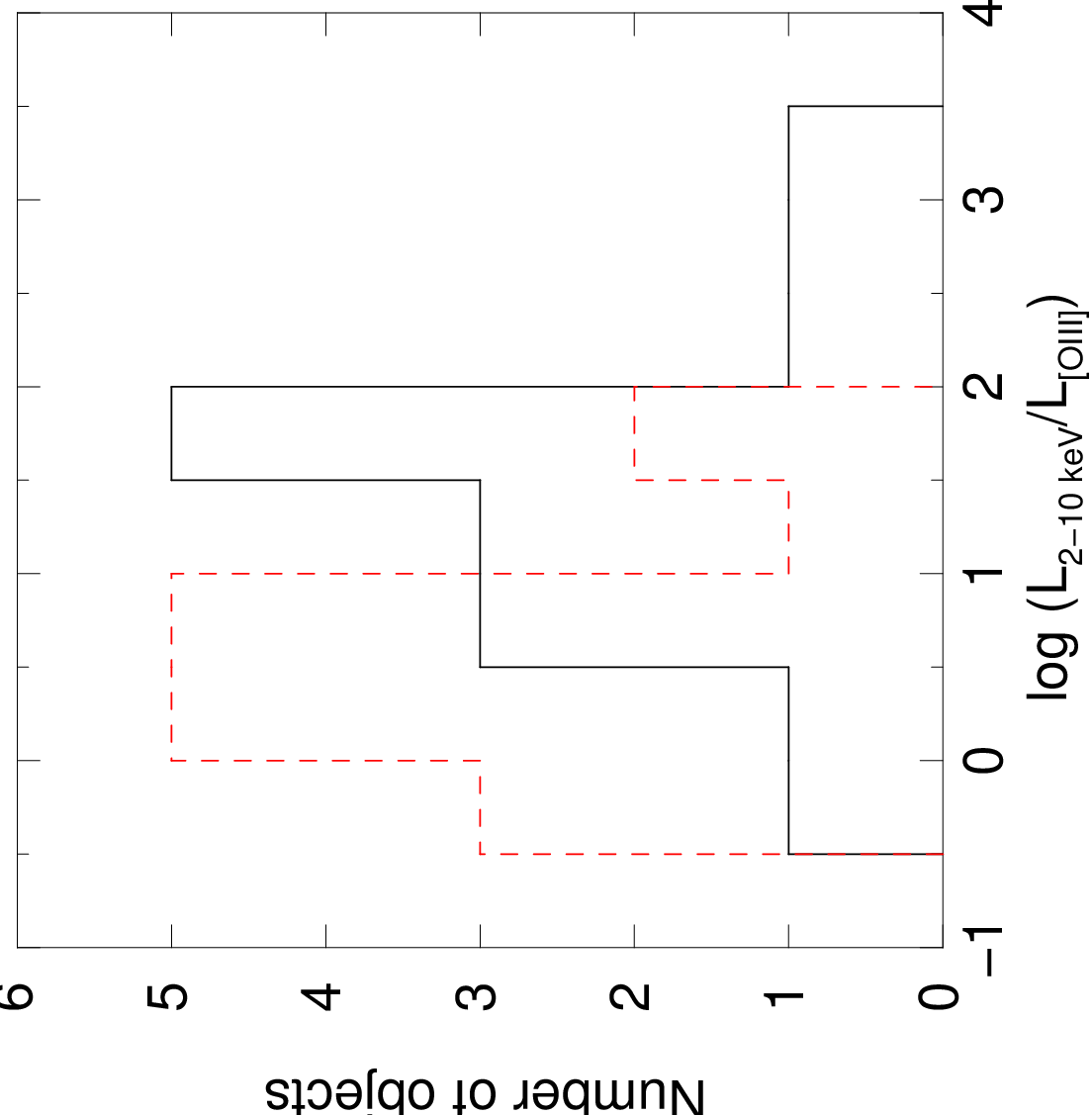}
\caption{Distribution of the ratio between intrinsic luminosity in
  the 2$-$10 keV band and reddening corrected [\ion{O}{3}] luminosity
  for our sample ({\it solid histogram}) and the Seyfert 2 sample
  compiled by Bassani et al. (1999) with {\it N}$_{\rm H}$ in the range
  of (0.6$-$20)$\times$10$^{23}$ cm$^{\rm -2}$ ({\it dashed histogram}).}
\label{figure:histogram_OIII_X_ratio}
\end{figure}

Figure \ref{figure:OIII} shows the correlation between the intrinsic
luminosities in the 2$-$10 keV band ({\it L}$_{\rm 2-10}^{\rm int}$)
and the reddening corrected [\ion{O}{3}] line luminosities ({\it
  L}$_{\rm [O\ III]}^{\rm int}$) for our sample and a large sample of
Seyfert 2 compiled by Bassani et al. (1999).  From the latter sample,
we used objects with $N_{\rm H}$ within (0.6$-$20)$\times$10$^{23}$
cm$^{-2}$, which is the range observed for our sample. Some objects
belong to both samples and such objects are regarded as members of our
sample.  The solid lines in Figure \ref{figure:OIII} correspond to
{\it L}$_{\rm 2-10}^{\rm int}$/{\it L}$_{\rm [O\ III]}^{\rm int}$ = 1,
10, and 100 from bottom to top. The distribution of the {\it L}$_{\rm
  2-10}^{\rm int}$/{\it L}$_{\rm [O\ III]}^{\rm int}$ ratios for the
two samples is shown in Figure \ref{figure:histogram_OIII_X_ratio}.
The ratios for Bassani's sample are in the range 1$-$100, while those
for most sources in our sample are $>$ 10. In particular, the ratios
for three sources (MCG+08$-$21$-$065, NGC 4138, and NGC 7172) are
greater than 100.

\begin{figure}[!t]
\centering
\includegraphics[angle=270,scale=.40]{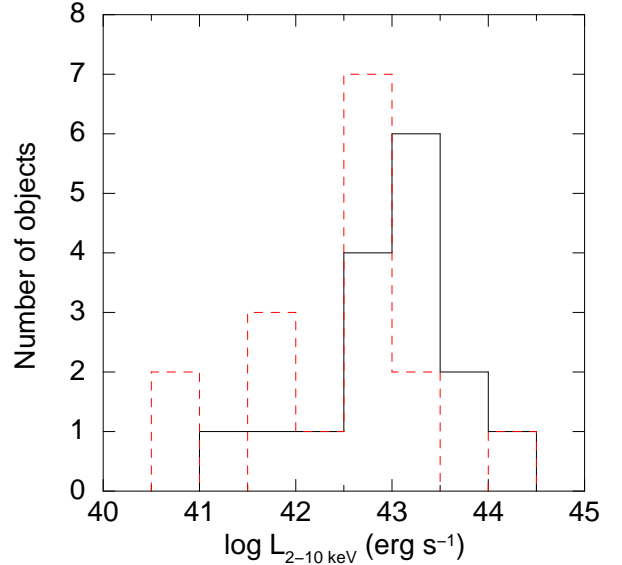}
\caption{Distribution of the intrinsic luminosity in the 2$-$10 keV band
  for our sample ({\it solid histogram}) and the Seyfert 2 sample
  compiled by Bassani et al. (1999) with {\it N}$_{\rm H}$ in the range
  of (0.6$-$20)$\times$10$^{23}$ cm$^{\rm -2}$ ({\it dashed histogram}).}
\label{figure:lumi_X}
\end{figure}

Netzer et al. (2006) showed that the ratio {\it L}$_{\rm 2-10}^{\rm
 int}$/{\it L}$_{\rm [O\ III]}^{\rm int}$ increases with {\it L}$_{\rm
 2-10}^{\rm int}$ such that 
 \begin{displaymath}
 {\rm log} \frac{{\it L}_{\rm 2-10}^{\rm int}}{{\it L}_{\rm [O\ III]}^{\rm int}} = (0.38\pm0.09){\rm log}{\it L}_{\rm 2-10}^{\rm int} - (15.0\pm4.0), 
 \end{displaymath}
using a sample obtained from Bassani et al. (1999) and supplemented by
Turner et al. (1997). Therefore, we compared the distribution of {\it
  L}$_{\rm 2-10}^{\rm int}$ between Bassani's and our samples to
examine whether the larger {\it L}$_{\rm 2-10}^{\rm int}$/{\it
  L}$_{\rm [O\ III]}^{\rm int}$ in our sample is explained by this
luminosity dependence. The distribution of {\it L}$_{\rm 2-10}^{\rm
  int}$ for Bassani's and our samples is shown in Figure
\ref{figure:lumi_X} as solid and dashed histograms, respectively. The
distribution for our sample is slightly biased toward higher
luminosities by log {\it L}$_{\rm 2-10}^{\rm int}$ $\sim$ 0.5.
According to Netzer's relation, this amount of shift in {\it L}$_{\rm
  2-10}^{\rm int}$ results in an increase of {\it L}$_{\rm 2-10}^{\rm
  int}$/{\it L}$_{\rm [O\ III]}^{\rm int}$ only by a factor of 1.5.
We attempted the two-sample Kolmogorov$-$Smirnov test to examine the
difference more quantitatively. The values of {\it L}$_{\rm 2-10}^{\rm
  int}$/{\it L}$_{\rm [O\ III]}^{\rm int}$ were scaled to an assumed
reference point {\it L}$_{\rm 2-10}^{\rm int}$ = 10$^{43}$ erg
s$^{-1}$ by using Netzer's relation, which was calculated by
 \begin{displaymath}
 {\rm log} \biggl[\frac{{\it L}_{\rm 2-10}^{\rm int}}{{\it L}_{\rm [O\ III]}^{\rm int}}\biggr]_{\rm scaled} =
  {\rm log} \frac{{\it L}_{\rm 2-10}^{\rm int}}{{\it L}_{\rm [O\ III]}^{\rm int}} - 0.38{\rm log}\biggl[\frac{{\it L}_{\rm 2-10}^{\rm int}}{10^{43} {\rm \ erg\ s}^{-1}}\biggr].
 \end{displaymath}
The Kolmogorov$-$Smirnov test showed that the distributions of the
scaled {\it L}$_{\rm 2-10}^{\rm int}$/{\it L}$_{\rm [O\ III]}^{\rm
  int}$ for Bassani's and our samples are drawn from the same parent
population with the probability of 9.3\%. Thus the difference of {\it
  L}$_{\rm 2-10}^{\rm int}$/{\it L}$_{\rm [O\ III]}^{\rm int}$ between
the two samples would not be explained solely by the luminosity
dependence.

We found that [\ion{O}{3}] luminosities for our sample are
intrinsically lower than those of Seyfert 2s studied so far at a given
X-ray luminosity, and agree with the above expectation. Although
luminosities of optical narrow emission lines are often utilized as a
good indicator of an intrinsic luminosity of an AGN and used for
constructing the most unbiased samples (e.g., Mulchaey et al. 1994;
Heckman 1995; Keel et al. 1994; Heckman et al. 2005), estimation of
intrinsic luminosities of an AGN based on [\ion{O}{3}] would have large
uncertainties and surveys that rely on [\ion{O}{3}] emission could be
subject to biases against buried AGNs. In order to obtain complete
unbiased samples of AGNs, hard X-ray surveys would be imperative.

\section{Conclusion}
We searched for AGNs, whose scattered emission is very weak, from the
2XMM Catalogue. In our selection procedure, we calculated HRs expected
for an object with a small scattering fraction using a model
consisting of absorbed and unabsorbed power laws and 22 sources were
selected as candidates from the 2XMM Catalogue. Spectral analysis was
conducted using the data observed with {\it XMM-Newton} for these 22
sources. Their X-ray spectra are represented by a combination of an
absorbed power law with a column density of $\sim$ 10$^{23-24}$
cm$^{-2}$, an unabsorbed power law, a narrow Gaussian line for the Fe
K emission, and some additional components. The photon indices of the
power-law components for 14 objects are in a typical range of Seyfert
2s ($\sim$ 1.5$-$2).  The photon indices for the others, which have
large uncertainties in $\Gamma$, were fixed at 1.9 in our analysis. We
found that the scattering fractions for our sample (except for NGC
4507 and IC 4995) were small compared to a typical value ($\sim$ 3\%)
of Seyfert 2s observed in the past. In addition, those of eight
sources are less than 0.5\%. If an opening angle of a torus is
responsible for the scattering fraction, objects in our sample would
be hidden by a geometrically thick torus with a small opening angle.

The ratios of {\it L}$_{\rm 0.5-2}^{\rm int}$/{\it L}$_{\rm FIR}$ for
about a half of our sample are in the range observed for starburst
galaxies. This result indicates that thermal emission from
a collisionally ionized plasma produced by starburst may contribute to
the soft X-ray emission, and true values of scattering fraction for
our sample would be smaller than the value calculated in this paper.

The distribution of {\it L}$_{\rm 2-10}^{\rm int}$/{\it L}$_{\rm
  [O\ III]}^{\rm int}$ for our sample is shifted to higher values than
that for other Seyfert 2s studied so far. {\it L}$_{\rm [O\ III]}^{\rm
  int}$ depends on the size of the NLR that is considered to be
existed in the opening part of the torus. Thus, this result also
indicates that the opening angle of the obscuring matter (or
scattering optical depth) for our sample is smaller than those for
other Seyfert 2, and there is a bias against such a type of AGN buried
in a very geometrically thick torus in surveys that rely on optical
emission.

\acknowledgments We thank an anonymous referee for useful comments
that improved the paper.  This paper is based on observations obtained
with {\it XMM-Newton}, an ESA science mission with instruments and
contributions directly funded by ESA Member States and the USA (NASA).
This research made use of the NASA/IPAC Extragalactic Database, which
is operated by the Jet Propulsion Laboratory, Calfornia Institute of
Technology, under contract with the National Aeronautics and Space
Administration. This work is supported by Grants-in-Aid for Scientific
Research 20740109 (Y.T.) and 21244017 (H.A.) from the Ministry of
Education, Culture, Sports, Science, and Technology of Japan.\\

Facilities: \facility{{\it XMM-Newton}}

\end{document}